\documentclass[reprint,amsmath,amssymb,aps,prl,longbibliography]{revtex4-1}
\usepackage{dcolumn}
\usepackage{color}
\usepackage{bm}      
\usepackage{xspace}
\usepackage{graphicx}

\newcommand{\be}{\begin{equation}}
\newcommand{\ee}{\end{equation}}
\newcommand{\bea}{\begin{eqnarray}}
\newcommand{\eea}{\end{eqnarray}}
\newcommand{\beaa}{\begin{eqnarray*}}
\newcommand{\eeaa}{\end{eqnarray*}}
\newcommand{\alphaI}{{$\alpha$-(BEDT-TTF)$_2$I$_3$}\xspace}

\begin{document}

\title{
Dynamical time-reversal and inversion symmetry breaking,
dimensional crossover, and chiral anomaly
in \alphaI
}

\author{Takao Morinari}
 \email{morinari.takao.5s@kyoto-u.ac.jp}
\affiliation{
  Graduate School of Human and Environmental Studies, 
  Kyoto University, Kyoto 606-8501, Japan
}

\date{\today}

\begin{abstract}
  In most Dirac semimetals, time-reversal and inversion symmetries
  are believed to play a crucial role in their stability.
We demonstrate that these symmetries are broken in Dirac fermions in the organic conductor \alphaI due to the strong electronic correlation.
The system is a three-dimensional type-II Dirac semimetal in the coherent inter-layer tunneling regime.
A chiral anomaly is predicted to be observed in the magnetoresistance when the magnetic field is tuned to the inter-layer tunneling direction.
Our result suggests that \alphaI is a useful platform to explore interplay between the chiral anomaly and the strong correlation and/or dimensionality.
\end{abstract}

\maketitle

Recently, the topological Dirac and Weyl semimetals
have attracted intense theoretical and experimental interest
because of their intriguing topological and electronic
properties \cite{Bansil2016,Armitage2018}.
In nodal semimetals, the conduction and valence bands
touch only at certain points in the Brillouin zone (BZ),
and the low-energy excitations are
described by a relativistic Dirac or Weyl equation,
where the velocity of light is replaced by the Fermi velocity \cite{CastroNeto2009}.
Two-dimensional (2D) Dirac semimetals
are realized in graphene \cite{Novoselov2005},
as is clearly demonstrated by the Dirac nature of the electronic transport
and the surface state of the three-dimensional (3D)
topological insulators \cite{Hasan2010,Qi2011},
where spin-orbit coupling plays a significant role.
Based on theoretical and experimental efforts,
3D Dirac semimetals are now realized experimentally,
for instance, in Na$_3$Bi \cite{Wang2012,Liu2014a}
and Cd$_3$As$_2$ \cite{Wang2013,Liu2014}.
Remarkably, the topological semimetals have deep connections
with particle physics because they provide solid state analogues
of relativistic chiral fermions \cite{Burkov2015,Armitage2018}
and lattice realizations \cite{Nielsen1983}
of the chiral anomaly of the quantum field theory \cite{Adler1969,Bell1969}.
Experimental evidence has been accumulated
about the existence of Fermi arc surface states \cite{Lv2015,Huang2015,Deng2016,Xu2016}
and the novel responses of the chiral anomaly to applied electronic and magnetic fields \cite{Huang2015a,Xiong2015,Li2015,Zhang2016,Hirschberger2016}.

Now it is well accepted that symmetries and spin-orbit coupling are the keys to realizing 3D Dirac semimetals in general.
Spin-orbit coupling can create a linear energy dispersion and a 3D Dirac semimetal appears when the symmetry conditions are met.
Since the net Chern numbers at each contact point are zero, we need additional crystal symmetries \cite{Young2012a,Wang2012,Wang2013,Manes2012,Yang2014}.
In such 3D Dirac semimetal systems,
Dirac points are on the symmetry lines in the BZ.
Our current understanding of the condition of the Dirac semimetal is mostly based on symmetry consideration and spin-orbit coupling,
though the latter is not necessary for some exceptional cases \cite{Song2018}.
Another important feature of Dirac and Weyl semimetals is
due to their lack of fundamental Lorentz symmetry.
In general, the energy dispersion around the contact points
is described by
${\varepsilon _ \pm }\left( {\bm{k}} \right)
= T\left( {\bm{k}} \right) \pm U\left( {\bm{k}} \right)$,
where $U\left( {\bm{k}} \right)$
describes the anisotropic cone
and the term $T\left( {\bm{k}} \right)$,
which is a linear function of ${\bm{k}}$,
describes the cone tilt.
In the case of a type-II semimetal,
the chiral anomaly only appears in the directions
where $T\left( {\bm{k}} \right) > U\left( {\bm{k}} \right)$ \cite{Soluyanov2015}
in contrast to a type-I semimetal
where $T\left( {\bm{k}} \right) < U\left( {\bm{k}} \right)$
in all directions.

In this Letter, we demonstrate that
the quasi-2D organic conductor, \alphaI,
is an unprecedented type of Dirac semimetal
that does not fit into our current understanding of Dirac semimetal conditions.
We show that contrary to other Dirac semimetals, 
both the time-reversal symmetry (TRS) and inversion symmetry are broken
and spin-orbit coupling plays no role.
In the symmetry broken state,
the energies of the Dirac points
are shifted asymmetrically from the Fermi energy 
and their positions are not symmetrically located
with respect to the origin of the BZ.
We also show that the system exhibits a dimensional crossover
from a type-I 2D Dirac semimetal
to a type-II 3D Dirac semimetal
upon entering the coherent inter-layer tunneling regime.

\begin{figure}[tbp]
  \includegraphics[width=\linewidth]{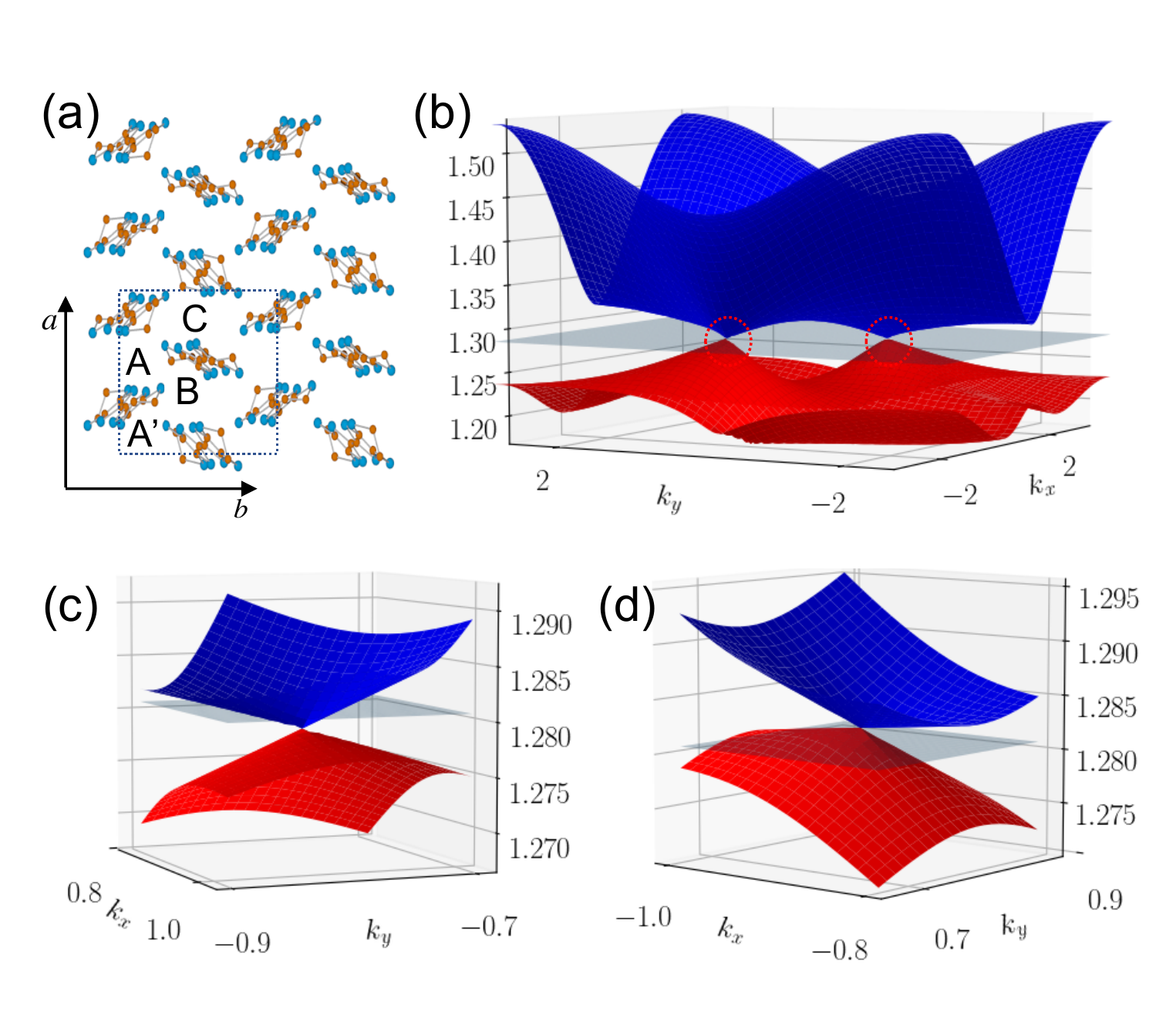}
\caption{
  \label{fig:ed2D:P0_8}   
  (a)Configuration of BEDT-TTF molecules 
  in a conducting plane of \alphaI.
  The rectangle shows a unit cell that contains
  four molecules, A, A$^{\prime}$, B, and C.
  The molecules stacked in the $a$ axis,
  which is taken as the $y$ axis
  and the $b$ axis is taken as the $x$ axis.
  (b) Conduction and valence bands are plotted as functions of
    $k_x$ and $k_y$ at $P=0.8$.
  There are two Dirac points at different energies,
  which are encircled by dotted circles,
    and they are located at the generic wave vectors.
    The horizontal plane denotes the Fermi energy.
    Magnified views of the Dirac nodes at
    ${\bm{k}}_D^{(1)}$ and
    ${\bm{k}}_D^{(2)}$ are shown in (c) and (d), respectively.
    The Dirac point at ${\bm{k}}_D^{(1)}$ is below the Fermi energy,
    while
    the Dirac point at ${\bm{k}}_D^{(2)}$ is above the Fermi energy.
}
\end{figure}

The organic conductor \alphaI with the space group $P\overline{1}$ 
is a 2D Dirac semimetal with a layered structure comprising of
Dirac fermion layers and insulating layers \cite{Kajita2014}.
The Dirac points are at non-high-symmetry points in the BZ
similar to certain inorganic materials with the $P\overline{1}$
space group \cite{Song2018,Zhang2019}.
In \alphaI, the unit cell is composed of
four BEDT-TTF molecules, where BEDT-TTF is bis(ethylenedithio)tetrathiafulvalene, A, A$^{\prime}$, B, and C, 
in the conduction layer \cite{Bender1984,Mori1984},
as shown in Fig.~\ref{fig:ed2D:P0_8}(a).
The system is metallic above 135~K but undergoes
a metal-insulator transition \cite{Kartsovnik85,Schwenk1985,Kajita1992}
at 135~K where a charge order stripe pattern forms
as confirmed by $^{13}$C-NMR (nuclear magnetic resonance) measurement \cite{Takahashi2003}.
Here, the short-range inter-site Coulomb repulsion
plays a key role \cite{Seo2000,Kino1996}.
The 2D Dirac semimetal appears when
the charge order is suppressed under high-pressure
as revealed by the tight-binding model calculation \cite{Katayama2006}
and confirmed by the first-principle calculations \cite{Ishibashi2006,Kino2006}.
It should be stressed that 
the band filling is fixed to 3/4 \cite{Mori1984}
and the Fermi energy is exactly at the Dirac point
within these calculations.
The inter-layer magnetoresistance, which is negative
and is in inversely proportional to the applied magnetic field,
clearly demonstrates the presence 
of the zero-energy Landau level \cite{Osada2008,Tajima2009,Morinari2009,Goerbig2008}.
Furthermore, the phase of the Dirac fermions is confirmed by
the Shubnikov-de Haas oscillation of the hole-doped sample,
where the sample is placed on polyethylene naphthalate substrate \cite{Tajima2013}.

Thus far, the presence of the massless Dirac fermion spectrum
has been established in \alphaI but the role of the strong electronic
correlation is unclear:
The charge order stripe pattern at ambient pressure
is replaced by charge disproportionation \cite{Moroto2004,Kobayashi2007},
where $n_{\textrm A} = n_{{\textrm A}^{\prime}}$
and $n_{\textrm B} \neq n_{\textrm C}$
with charge density at molecule $\alpha$ denoted by $n_\alpha$.
The on-site Coulomb repulsion plays a central role
in the determination of the charge densities
but the inter-site Coulomb interaction plays a minor role.
The purpose of this study is to demonstrate that
the inter-site Coulomb interaction plays a crucial role
in stabilizing a non-trivial Dirac semimetal state.

The Hamiltonian of electrons within a conduction layer
is given by
${\mathcal H} = {{\mathcal H}_0} + {\mathcal H}_{\textrm{int}}$.
The first term describes the hopping between
the molecules and in the momentum space,
\be
{{\mathcal H}_0} = \sum\limits_{\bm{k},\sigma=\uparrow,\downarrow} 
{c_{\bm{k}\sigma}^\dag {{\mathcal H}^0_{\bm{k}}}{c_{\bm{k}\sigma}}},
\ee
where 
$c_{\bm{k}\sigma}^\dag  = \left( 
{c_{{\bm{k}}1\sigma}^\dag }, {c_{{\bm{k}}2\sigma}^\dag }, 
{c_{{\bm{k}}3\sigma}^\dag }, {c_{{\bm{k}}4\sigma}^\dag }
\right)$
is the four-component creation operator
for an electron with momentum $\bm{k}$
and spin component $\sigma$.
The indices 1, 2, 3, and 4 represent molecules
A, A$^{\prime}$, B, and C, respectively.
The matrix elements of $H^0_{\bm{k}}$ are
${\left( {{\mathcal H}_{\bm{k}}^0} \right)_{12}} 
= {t_{a3}}{e^{ - i{\bm{k}} \cdot {{\bm{d}}_1}}} 
+ {t_{a2}}{e^{i{\bm{k}} \cdot {{\bm{d}}_1}}}$,
${\left( {{\mathcal H}_{\bm{k}}^0} \right)_{13}} 
= {t_{b3}}
{e^{ - i{\bm{k}} \cdot {{\bm{d}}_3}}} 
+ {t_{b2}}
{e^{i{\bm{k}} \cdot {{\bm{d}}_2}}}$,
${\left( {{\mathcal H}_{\bm{k}}^0} \right)_{14}} 
= {t_{b4}}
{e^{ - i{\bm{k}} \cdot {{\bm{d}}_2}}} 
+ 
{t_{b1}}
{e^{i{\bm{k}} \cdot {{\bm{d}}_3}}}$,
${\left( {{\mathcal H}_{\bm{k}}^0} \right)_{23}} 
=
{t_{b2}}
{e^{ - i{\bm{k}} \cdot {{\bm{d}}_2}}} 
+ 
{t_{b3}}
{e^{i{\bm{k}} \cdot {{\bm{d}}_3}}}$,
${\left( {{\mathcal H}_{\bm{k}}^0} \right)_{24}} 
=
{t_{b1}}
{e^{ - i{\bm{k}} \cdot {{\bm{d}}_3}}} 
+ 
{t_{b4}}
{e^{i{\bm{k}} \cdot {{\bm{d}}_2}}}$, and
${\left( {{\mathcal H}_{\bm{k}}^0} \right)_{34}} 
=
{t_{a1}}{e^{i{\bm{k}} \cdot {{\bm{d}}_1}}} 
+ {t_{a1}}{e^{ - i{\bm{k}} \cdot {{\bm{d}}_1}}}$,
where the displacement vectors, ${\bm d}_1$, 
${\bm d}_2$, and ${\bm d}_3$, respectively, are defined by
${{\bm{d}}_1} = \left( 0, a/2\right)$,
${{\bm{d}}_2} = \left( b/2, -a/4 \right)$, 
and ${{\bm{d}}_3} = \left( b/2, a/4\right)$ \cite{Kajita2014}.
Hereafter, we measure the energies in units of eV and we set $a=1$ and $b=1$ for the lattice constants.
The transfer energies are pressure dependent
and given by \cite{Kobayashi2007}
${t_\alpha } = {C_\alpha }\left( {1 + {b_\alpha }P} \right)$.
Here, the pressure $P$ is in units of GPa.
The numerical coefficients $C_\alpha$ are
$C_{a1}=-0.028$, $C_{a2}=0.048$, and $C_{a3}=-0.020$
for the stacking direction and 
$C_{b1}=0.123$, $C_{b2}=0.140$, $C_{b3}=-0.062$,
and $C_{b4}=-0.025$ for the other directions.
The numerical constants $b_\alpha$ are
$b_{a1}=0.89$, $b_{a2}=1.67$, $b_{a3}=-0.25$, $b_{b1}=0$,
$b_{b2}=0.11$, $b_{b3}=0.32$, and $b_{b4}=0$.

The interaction term ${\mathcal H}_{\textrm{int}}$
describes a strong electronic correlation
that is given by
\be
{\mathcal H}_{\textrm{int}}
= U\sum\limits_{j,\alpha } {{n_{j\alpha  \uparrow }}{n_{j\alpha  \downarrow }}}
+ \sum\limits_{i,j,\alpha,\beta}
{{V_{i\alpha, j\beta }}{n_{i\alpha }}{n_{j\beta }}}.
\ee
Here, 
the charge density of the electrons with spin $\sigma$ 
at molecule $\alpha$ in site $j$
is denoted by $n_{j\alpha \sigma}$,
and we define $n_{j\alpha} = n_{j\alpha\uparrow} + n_{j\alpha\downarrow}$.
The first term in the right-hand side describes the on-site Coulomb interaction
while the second term describes the Coulomb interaction between nearest neighbor molecules.
We set $V_{i1,j2}=V_{i3,j4}=V_c$ along 
the stacking direction of the BEDT-TTF molecules (the $a$-axis)
and $V_{i\alpha,j\beta} = V_p$ otherwise.
These interactions lead to an insulating state,
which is a charge ordered state \cite{Kino1996,Seo2000},
under ambient pressure
and lead to dynamical TRS and inversion symmetry
breaking in the Dirac semimetal phase
as we illustrate below.

It has not yet been considered explicitly,
except for in limited cases \cite{Sasaki2014,Morinari2014},
but the significant electronic correlation
in the Dirac semimetal state,
in the presence of
the inter-molecule interaction $V_{\alpha\beta}$,
is the bond correlation
described by 
\be
{\chi _{\alpha \sigma ,\beta \sigma ', \pm }} 
= \frac{1}{N}\sum\limits_{\bm{k}} {{e^{ - i{\bm{k}} 
\cdot {\bm{d}}_{\alpha \beta }^{\left(  \pm  \right)}}}
\left\langle {c_{{\bm{k}}\alpha \sigma }^\dagger 
{c_{{\bm{k}}\beta \sigma '}}} \right\rangle },
\ee
where $N$ is the number of BZ points
and
${\bm{d}}_{13}^{\left(  +  \right)} = {{\bm{d}}_2}$,
${\bm{d}}_{13}^{\left(  -  \right)} = {{\bm{d}}_3}$, etc.
Non-zero values of 
${\chi _{\alpha \sigma ,\beta \sigma ', \pm }}$
can break the TRS and inversion symmetry.
We also include the site order
defined as
\be
{{n_{\alpha \sigma }}}
= \frac{1}{N}\sum\limits_{\bm{k}} {\left\langle {c_{{\bm{k}}\alpha \sigma }^\dagger 
{c_{{\bm{k}}\alpha \sigma }}} \right\rangle },
\ee
which exhibits charge disproportionation
in the Dirac semimetal \cite{Moroto2004,Kobayashi2007}.
The resulting mean field Hamiltonian is denoted by
\be
  {{\mathcal H}_{\textrm{mf}}} = \sum\limits_{\bm{k},\sigma,\sigma'} {c_{\bm{k}\sigma}^\dag
    \left[
      {\mathcal H}\left( {\bm{k}} \right)
      \right]_{\sigma \sigma'}
      {c_{\bm{k}\sigma'}}}.
  \ee
  The matrix ${\mathcal H}\left( {\bm{k}} \right)$ is
  $8 \times 8$.
  We solve the self-consistent equations
  for ${\chi _{\alpha \sigma ,\beta \sigma ', \pm }}$
  and ${{n_{\alpha \sigma }}}$.

  Now we describe symmetries of the system.
To describe the symmetry operations, 
we define
\be
  {{\mathcal X}_{\mu \nu \lambda }}
  = {\sigma _\mu } \otimes {\sigma _\nu } \otimes {\sigma _\lambda },
  \ee
  where $\mu, \nu, \lambda = 0,1,2,3$.
  Here, $\sigma_1$, $\sigma_2$, and $\sigma_3$
  denote the Pauli matrices and
  $\sigma _0$ is the $2 \times 2$ unit matrix.
  In the definition of ${{\mathcal X}_{\mu \nu \lambda }}$,
  the first two Pauli matrices
  act on the four molecule indices
  and the last Pauli matrix acts on the spin index.
The TRS is
\be
  {\mathcal X}_{002} {{\mathcal H}^*}\left( { - {\bm{k}}} \right)
  {\mathcal X}_{002}
= {\mathcal H}\left( {\bm{k}} \right).
\ee
The system has an inversion center \cite{Mori1984,AsanoHotta2011,Piechon2013}
between A and A$^{\prime}$.
The symmetry operation associated with this inversion is
\be
{\mathcal I} {{\mathcal H}}\left( { - {\bm{k}}} \right) {\mathcal I}
= {\mathcal H}\left( {\bm{k}} \right),
\ee
where
${{\mathcal I}} = i\left(
  {{{\mathcal X}_{103}} + {{\mathcal X}_{133}}
    + {{\mathcal X}_{003}} - {{\mathcal X}_{033}}} \right)/2$.
In the absence of the interactions,
we see that the system is invariant under these symmetry operations.
The situation does not change if we include the on-site Coulomb repulsion.
If we restrict the self-consistent calculation to
the Hartree level,
these symmetries are unbroken
but the exchange correlation associated with the inter-site
Coulomb repulsion leads to breaking of {\textit{both}} symmetries.

We present the energy dispersion
of the conduction and valence bands in Fig.~\ref{fig:ed2D:P0_8}(b), (c), and (d).
The conduction and valence bands are both spin degenerate.
Contrary to the naive expectation,
there is {\textit{no mass gap}}
in the Dirac fermion spectrum.
We find that the two Dirac points
are at
${{\bm{k}}_D^{(1)} } = \left( {0.9250}, - 0.7978 \right)$
where energy ${\varepsilon _D^{(1)} } = 1.2788$
and at 
${{\bm{k}}_D^{(2)} } = \left( - 0.8988,0.7631 \right)$
where energy ${\varepsilon _D^{(2)} } = {1.2822}$.
Here, the Fermi energy is $\varepsilon_F=1.2807$ and
we note that
${\bm{k}}_D^{\left( 1 \right)} \ne  - {\bm{k}}_D^{\left( 2 \right)}$
and
$\varepsilon _D^{\left( 1 \right)} < {\varepsilon _F}
< \varepsilon _D^{\left( 2 \right)}$.
By changing the pressure, the Dirac points move
in the BZ and the electronic correlation changes
as well \footnote{See Supplemental Material for details about
  the move of the Dirac points in the BZ and the change
  of the electronic correlation.}.
Hereafter, we take $U=0.4$, $V_c=0.17$, $V_p=0.05$, and $P=0.8$.
This set of interaction parameters reproduces the experimentally
observed stripe pattern in the insulating state \cite{Kobayashi2007}.

Although the renormalized hopping parameters
break the inversion symmetry,
the charges do not.
In fact, we find that $n_{\textrm A}=n_{{\textrm A}^{\prime}}=1.4554$
where $n_{\textrm B}=1.2204$ and $n_{\textrm C}=1.8696$.
This symmetry is protected by
the strong correlation associated with $U$,
whereas $V_c$ favors breaking this symmetry.
The system undergoes a quantum phase transition
as we increase the value of $V_c$,
merging the two Dirac points
\footnote{See Supplemental Material for details about
  symmetry breaking}.

\begin{figure}[tbp]
  \includegraphics[width=\linewidth]{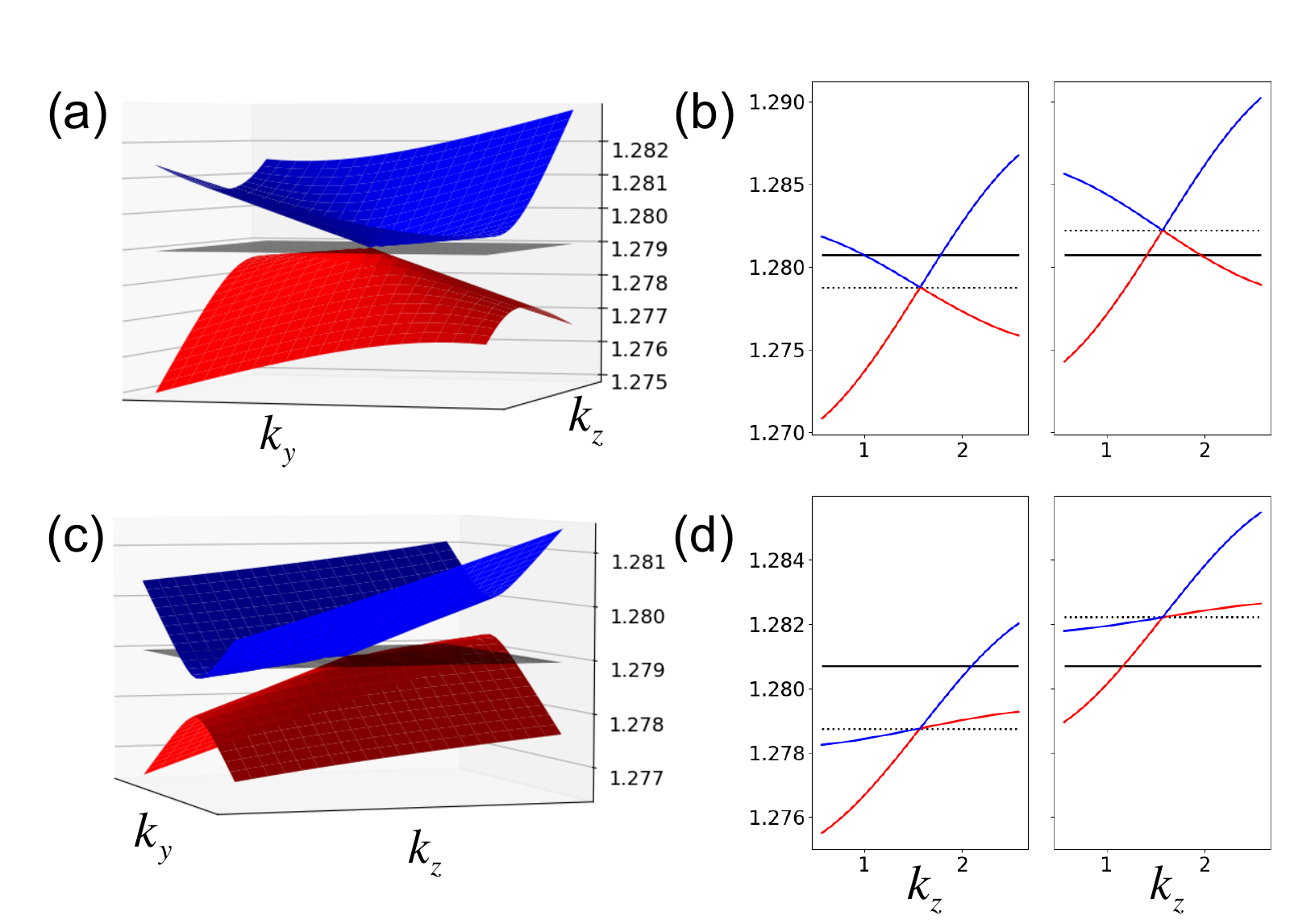}  
\caption{
  \label{fig:DP2dEDP3d} 
  Type-I Dirac node for $t_2/t_1=4$ (a)
  and its energy dispersion along the $k_z$ axis (b).
  Type-II Dirac node for $t_2/t_1=1$ (c)
  and its energy dispersion along the $k_z$ axis (d).  
  In both cases, we set $t_1=0.001$.
  The horizontal planes in (a) and (c) denote
  the energies of the Dirac points.
  In (b) and (d), the Fermi energy is denoted by
  the solid lines
  and the Dirac point energies
  are denoted by the dotted lines.
}
\end{figure}

Since \alphaI has a layered structure and
high-mobility \cite{Kajita2014}
$\sim 10^5 {\textrm{cm}^2} {\textrm V}^{-1} {\textrm s}^{-1}$,
the system undergoes a dimensional crossover
from the 2D electronic state to the 3D electronic state.
When the interlayer tunneling is incoherent,
the electronic structure is 2D and
when
the interlayer tunneling becomes coherent \cite{McKenzie1998},
the electronic structure is 3D.
Due to the high-mobility value of \alphaI,
the crossover temperature is in the order of
the inter-layer tunneling amplitude.
We describe the inter-layer tunneling
between adjacent layers
with the matrix $t_1 {\mathcal X}_{000} + t_2 {\mathcal X}_{010}$,
where the first term is the tunneling between the same molecules.
The second term is the tunneling between ${\textrm A}$ and ${\textrm A}^{\prime}$
and between ${\textrm B}$ and ${\textrm C}$, where
these pairs of molecules are aligned in the stacking direction.
We note that these terms do not break
the TRS or
inversion symmetry, ${\mathcal I}$.
Therefore, the contact points are stable against them.
Because of the mirror reflection about the $a$-$b$ plane,
two copies of each Dirac point appear
at $k_z=\pm \pi/2$.
Since interlayer hopping parameters $t_1$ and $t_2$
are much smaller than the intralayer parameters,
we include their effects based on the 2D result.
We observe that the Dirac cone is tilted along the $k_z$ direction by $t_1$
as shown in Fig.~\ref{fig:DP2dEDP3d}.
For $t_1/t_2 > \eta_1$ (Fig.~\ref{fig:DP2dEDP3d}(c) and (d)),
both Dirac cones are type-II \cite{Soluyanov2015},
which is to be applied to \alphaI,
while for $t_1/t_2 < \eta_2$ (Fig.~\ref{fig:DP2dEDP3d}(a) and (b)),
both Dirac cones are type-I.
Here, $\eta_1=0.6867$
and $\eta_2=0.6827$.
Interestingly, for $\eta_2 < t_1/t_2 < \eta_1$,
the Dirac cones at ${\bm k}_D^{(1)}$ with $k_z=\pm \pi/2$
are type-I
and the Dirac cones at ${\bm k}_D^{(2)}$ with $k_z=\pm \pi/2$
are type-II,
where we can expect the partial chiral anomaly effect to be associated
with the type-I Dirac cones.

Now we consider the chiral anomaly in this system.
When a magnetic field $B_z$ is applied along the $z$-direction,
the spectrum of the Landau levels is given by
\be
\varepsilon _{n,{k_z}}^{ \pm ,\tau } =  - 2{t_1}\cos {k_z}
\pm \sqrt {\varepsilon _{n\tau }^2 + 4\eta _\tau ^2t_2^2{{\cos }^2}{k_z}},
\ee
where ${\varepsilon _{n\tau }}$ is the 2D Landau level
at the Dirac point ${\bm{k}}_D^{(\tau)}$
($\tau=1,2$) given by \cite{Morinari2009,Goerbig2008}
\be
   {\varepsilon _{n\tau }} = {\left( {1 - \lambda _\tau ^2} \right)^{3/4}}
   \frac{{\hbar v_\tau ^{{\textrm{2D}}}}}{{{\ell_z}}}\sqrt {2|n|}.
   \ee
   Here, $n$ is an integer,
   $\lambda _\tau$ is the tilt parameter of the Dirac cone,
   $v_\tau ^{\textrm{2D}}$ is the averaged Fermi velocity,
   and $\ell_z=\sqrt{\hbar/|eB_z|}$ is the magnetic length,
   where $\hbar$ is the reduced Planck constant and
   $e$ is the electron charge.
   In \alphaI, $\lambda _\tau$ and $v_\tau ^{\textrm{2D}}$
   can be estimated experimentally
   from the analysis of the interlayer magnetoresistance.
   It is found that 
   $\left(1-\lambda_\tau^2 \right)^{3/4}
   v_\tau ^{\textrm{2D}} \simeq 5 \times 10^{-4}$~m/s \cite{Sugawara2010mr},
   where
   $\sqrt{1-\lambda _\tau^2} \simeq 0.05$ \cite{Tajima2018}.

   Since the system is a 3D Dirac semimetal,
   the $n=0$ Landau level has a chiral mode,
   $\varepsilon _{0,\pm \pi /2 + \delta {k_z}}^{ \pm ,\tau }
   = \pm \hbar v_z^\tau \delta {k_z}$,
   such that
   \be
   v_z^\tau  =  \frac{2a_z}{\hbar}\left( {{t_1} - {\eta _\tau }{t_2}} \right),
   \ee
   where the lattice constant $a_z$ is explicitly shown.
   The plus (minus) sign is for the Dirac cone at $k_z=\pi/2 (-\pi/2)$.
   Because of the current flow between the two Dirac nodes,
   the negative magnetoresistance
   is observed when the magnetic field is tuned,
   within the angle $\delta \theta$,
   to the direction of the inter-layer hopping,
   which is taken as the $z$-axis here for simplicity.
   We emphasize that the effect is limited to
   the magnetic field directions close
   to the inter-layer tunneling direction
   because of the type-II nature of the chiral anomaly \cite{Soluyanov2015}.
   We note that $\delta \theta$ is approximately proportional to $1/t_1$,
   and we find $\delta \theta \simeq 0.36^{\circ}$
   when $t_1=0.001$.

The calculation above can be extended to
include the effect of spin-orbit coupling \cite{Winter2017,Osada2018}.
Due to the configuration of the molecules in the unit cell,
there is no spin-orbit coupling along the molecule stacking direction.
Although there is some ambiguity in the choice of the
spin-orbit coupling parameters,
a clear conclusion can be reached that
the spin-orbit coupling does not create any mass gap
at the Dirac points.
Here, the exchange correlation effect also plays a crucial role
\footnote{See Supplemental Material for details about
  the effect of spin-orbit coupling.}.

In conclusion, we have demonstrated that \alphaI is a new type of Dirac semimetal with remarkable features.
Contrary to other Dirac semimetals, both the TRS and inversion are broken in a non-trivial way.
This result clearly extends our current understanding of the symmetry condition for Dirac semimetals.
In particular, our new Dirac semimetal can be used to deepen our understanding of chiral anomaly;
we expect chiral anomaly to exist in 3D but not in 2D and
the transition between them can be investigated in \alphaI
through the negative magnetoresistance in the direction
of the inter-layer tunneling.
One limitation is that the sample must be in a pressure cell,
though the pressure is useful to control the electronic correlation of the system and investigate the interplay
between chiral anomaly and other electronic states.

\noindent
\textbf{Acknowledgments.}
  The author thanks N. Tajima for participating in helpful discussions and
  sharing experimental data.
  This work was supported by JSPS KAKENHI Grant Number JP18K18739.

\bibliography{../../../../refs/lib}

\begin{thebibliography}{60}%
\makeatletter
\providecommand \@ifxundefined [1]{%
 \@ifx{#1\undefined}
}%
\providecommand \@ifnum [1]{%
 \ifnum #1\expandafter \@firstoftwo
 \else \expandafter \@secondoftwo
 \fi
}%
\providecommand \@ifx [1]{%
 \ifx #1\expandafter \@firstoftwo
 \else \expandafter \@secondoftwo
 \fi
}%
\providecommand \natexlab [1]{#1}%
\providecommand \enquote  [1]{``#1''}%
\providecommand \bibnamefont  [1]{#1}%
\providecommand \bibfnamefont [1]{#1}%
\providecommand \citenamefont [1]{#1}%
\providecommand \href@noop [0]{\@secondoftwo}%
\providecommand \href [0]{\begingroup \@sanitize@url \@href}%
\providecommand \@href[1]{\@@startlink{#1}\@@href}%
\providecommand \@@href[1]{\endgroup#1\@@endlink}%
\providecommand \@sanitize@url [0]{\catcode `\\12\catcode `\$12\catcode
  `\&12\catcode `\#12\catcode `\^12\catcode `\_12\catcode `\%12\relax}%
\providecommand \@@startlink[1]{}%
\providecommand \@@endlink[0]{}%
\providecommand \url  [0]{\begingroup\@sanitize@url \@url }%
\providecommand \@url [1]{\endgroup\@href {#1}{\urlprefix }}%
\providecommand \urlprefix  [0]{URL }%
\providecommand \Eprint [0]{\href }%
\providecommand \doibase [0]{http://dx.doi.org/}%
\providecommand \selectlanguage [0]{\@gobble}%
\providecommand \bibinfo  [0]{\@secondoftwo}%
\providecommand \bibfield  [0]{\@secondoftwo}%
\providecommand \translation [1]{[#1]}%
\providecommand \BibitemOpen [0]{}%
\providecommand \bibitemStop [0]{}%
\providecommand \bibitemNoStop [0]{.\EOS\space}%
\providecommand \EOS [0]{\spacefactor3000\relax}%
\providecommand \BibitemShut  [1]{\csname bibitem#1\endcsname}%
\let\auto@bib@innerbib\@empty
\bibitem [{\citenamefont {Bansil}\ \emph {et~al.}(2016)\citenamefont {Bansil},
  \citenamefont {Lin},\ and\ \citenamefont {Das}}]{Bansil2016}%
  \BibitemOpen
  \bibfield  {author} {\bibinfo {author} {\bibfnamefont {A.}~\bibnamefont
  {Bansil}}, \bibinfo {author} {\bibfnamefont {Hsin}\ \bibnamefont {Lin}}, \
  and\ \bibinfo {author} {\bibfnamefont {Tanmoy}\ \bibnamefont {Das}},\
  }\bibfield  {title} {\enquote {\bibinfo {title} {Colloquium: Topological band
  theory},}\ }\href {\doibase 10.1103/revmodphys.88.021004} {\bibfield
  {journal} {\bibinfo  {journal} {Rev. Mod. Phys.}\ }\textbf {\bibinfo {volume}
  {88}},\ \bibinfo {pages} {021004} (\bibinfo {year} {2016})}\BibitemShut
  {NoStop}%
\bibitem [{\citenamefont {Armitage}\ \emph {et~al.}(2018)\citenamefont
  {Armitage}, \citenamefont {Mele},\ and\ \citenamefont
  {Vishwanath}}]{Armitage2018}%
  \BibitemOpen
  \bibfield  {author} {\bibinfo {author} {\bibfnamefont
  {N.{\hspace{0.167em}}P.}\ \bibnamefont {Armitage}}, \bibinfo {author}
  {\bibfnamefont {E.{\hspace{0.167em}}J.}\ \bibnamefont {Mele}}, \ and\
  \bibinfo {author} {\bibfnamefont {Ashvin}\ \bibnamefont {Vishwanath}},\
  }\bibfield  {title} {\enquote {\bibinfo {title} {{Weyl and Dirac semimetals
  in three-dimensional solids}},}\ }\href {\doibase
  10.1103/revmodphys.90.015001} {\bibfield  {journal} {\bibinfo  {journal}
  {Rev. Mod. Phys.}\ }\textbf {\bibinfo {volume} {90}},\ \bibinfo {pages}
  {015001} (\bibinfo {year} {2018})}\BibitemShut {NoStop}%
\bibitem [{\citenamefont {Castro~Neto}\ \emph {et~al.}(2009)\citenamefont
  {Castro~Neto}, \citenamefont {Guinea}, \citenamefont {Peres}, \citenamefont
  {Novoselov},\ and\ \citenamefont {Geim}}]{CastroNeto2009}%
  \BibitemOpen
  \bibfield  {author} {\bibinfo {author} {\bibfnamefont {A.~H.}\ \bibnamefont
  {Castro~Neto}}, \bibinfo {author} {\bibfnamefont {F.}~\bibnamefont {Guinea}},
  \bibinfo {author} {\bibfnamefont {N.~M.~R.}\ \bibnamefont {Peres}}, \bibinfo
  {author} {\bibfnamefont {K.~S.}\ \bibnamefont {Novoselov}}, \ and\ \bibinfo
  {author} {\bibfnamefont {A.~K.}\ \bibnamefont {Geim}},\ }\bibfield  {title}
  {\enquote {\bibinfo {title} {The electronic properties of graphene},}\ }\href
  {\doibase 10.1103/RevModPhys.81.109} {\bibfield  {journal} {\bibinfo
  {journal} {Rev. Mod. Phys.}\ }\textbf {\bibinfo {volume} {81}},\ \bibinfo
  {pages} {109--162} (\bibinfo {year} {2009})}\BibitemShut {NoStop}%
\bibitem [{\citenamefont {Novoselov}\ \emph {et~al.}(2005)\citenamefont
  {Novoselov}, \citenamefont {Geim}, \citenamefont {Morozov}, \citenamefont
  {Jiang}, \citenamefont {Katsnelson}, \citenamefont {Grigorieva},
  \citenamefont {Dubonos},\ and\ \citenamefont {Firsov}}]{Novoselov2005}%
  \BibitemOpen
  \bibfield  {author} {\bibinfo {author} {\bibfnamefont {K.~S.}\ \bibnamefont
  {Novoselov}}, \bibinfo {author} {\bibfnamefont {A.~K.}\ \bibnamefont {Geim}},
  \bibinfo {author} {\bibfnamefont {S.~V.}\ \bibnamefont {Morozov}}, \bibinfo
  {author} {\bibfnamefont {D.}~\bibnamefont {Jiang}}, \bibinfo {author}
  {\bibfnamefont {M.~I.}\ \bibnamefont {Katsnelson}}, \bibinfo {author}
  {\bibfnamefont {I.~V.}\ \bibnamefont {Grigorieva}}, \bibinfo {author}
  {\bibfnamefont {S.~V.}\ \bibnamefont {Dubonos}}, \ and\ \bibinfo {author}
  {\bibfnamefont {A.~A.}\ \bibnamefont {Firsov}},\ }\bibfield  {title}
  {\enquote {\bibinfo {title} {{Two-dimensional gas of massless Dirac fermions
  in graphene}},}\ }\href@noop {} {\bibfield  {journal} {\bibinfo  {journal}
  {Nature}\ }\textbf {\bibinfo {volume} {438}},\ \bibinfo {pages} {197}
  (\bibinfo {year} {2005})}\BibitemShut {NoStop}%
\bibitem [{\citenamefont {Hasan}\ and\ \citenamefont {Kane}(2010)}]{Hasan2010}%
  \BibitemOpen
  \bibfield  {author} {\bibinfo {author} {\bibfnamefont {M.~Z.}\ \bibnamefont
  {Hasan}}\ and\ \bibinfo {author} {\bibfnamefont {C.~L.}\ \bibnamefont
  {Kane}},\ }\bibfield  {title} {\enquote {\bibinfo {title} {Colloquium:
  Topological insulators},}\ }\href {\doibase 10.1103/revmodphys.82.3045}
  {\bibfield  {journal} {\bibinfo  {journal} {Rev. Mod. Phys.}\ }\textbf
  {\bibinfo {volume} {82}},\ \bibinfo {pages} {3045--3067} (\bibinfo {year}
  {2010})}\BibitemShut {NoStop}%
\bibitem [{\citenamefont {Qi}\ and\ \citenamefont {Zhang}(2011)}]{Qi2011}%
  \BibitemOpen
  \bibfield  {author} {\bibinfo {author} {\bibfnamefont {Xiao-Liang}\
  \bibnamefont {Qi}}\ and\ \bibinfo {author} {\bibfnamefont {Shou-Cheng}\
  \bibnamefont {Zhang}},\ }\bibfield  {title} {\enquote {\bibinfo {title}
  {Topological insulators and superconductors},}\ }\href {\doibase
  10.1103/revmodphys.83.1057} {\bibfield  {journal} {\bibinfo  {journal} {Rev.
  Mod. Phys.}\ }\textbf {\bibinfo {volume} {83}},\ \bibinfo {pages}
  {1057--1110} (\bibinfo {year} {2011})}\BibitemShut {NoStop}%
\bibitem [{\citenamefont {Wang}\ \emph {et~al.}(2012)\citenamefont {Wang},
  \citenamefont {Sun}, \citenamefont {Chen}, \citenamefont {Franchini},
  \citenamefont {Xu}, \citenamefont {Weng}, \citenamefont {Dai},\ and\
  \citenamefont {Fang}}]{Wang2012}%
  \BibitemOpen
  \bibfield  {author} {\bibinfo {author} {\bibfnamefont {Zhijun}\ \bibnamefont
  {Wang}}, \bibinfo {author} {\bibfnamefont {Yan}\ \bibnamefont {Sun}},
  \bibinfo {author} {\bibfnamefont {Xing-Qiu}\ \bibnamefont {Chen}}, \bibinfo
  {author} {\bibfnamefont {Cesare}\ \bibnamefont {Franchini}}, \bibinfo
  {author} {\bibfnamefont {Gang}\ \bibnamefont {Xu}}, \bibinfo {author}
  {\bibfnamefont {Hongming}\ \bibnamefont {Weng}}, \bibinfo {author}
  {\bibfnamefont {Xi}~\bibnamefont {Dai}}, \ and\ \bibinfo {author}
  {\bibfnamefont {Zhong}\ \bibnamefont {Fang}},\ }\bibfield  {title} {\enquote
  {\bibinfo {title} {Dirac semimetal and topological phase transitions in
  ${{A}}_{3}${Bi} (${A}=\text{{Na}}$, {K}, {Rb})},}\ }\href {\doibase
  10.1103/PhysRevB.85.195320} {\bibfield  {journal} {\bibinfo  {journal} {Phys.
  Rev. B}\ }\textbf {\bibinfo {volume} {85}},\ \bibinfo {pages} {195320}
  (\bibinfo {year} {2012})}\BibitemShut {NoStop}%
\bibitem [{\citenamefont {Liu}\ \emph {et~al.}(2014{\natexlab{a}})\citenamefont
  {Liu}, \citenamefont {Zhou}, \citenamefont {Zhang}, \citenamefont {Wang},
  \citenamefont {Weng}, \citenamefont {Prabhakaran}, \citenamefont {Mo},
  \citenamefont {Shen}, \citenamefont {Fang}, \citenamefont {Dai},
  \citenamefont {Hussain},\ and\ \citenamefont {Chen}}]{Liu2014a}%
  \BibitemOpen
  \bibfield  {author} {\bibinfo {author} {\bibfnamefont {Z.~K.}\ \bibnamefont
  {Liu}}, \bibinfo {author} {\bibfnamefont {B.}~\bibnamefont {Zhou}}, \bibinfo
  {author} {\bibfnamefont {Y.}~\bibnamefont {Zhang}}, \bibinfo {author}
  {\bibfnamefont {Z.~J.}\ \bibnamefont {Wang}}, \bibinfo {author}
  {\bibfnamefont {H.~M.}\ \bibnamefont {Weng}}, \bibinfo {author}
  {\bibfnamefont {D.}~\bibnamefont {Prabhakaran}}, \bibinfo {author}
  {\bibfnamefont {S.-K.}\ \bibnamefont {Mo}}, \bibinfo {author} {\bibfnamefont
  {Z.~X.}\ \bibnamefont {Shen}}, \bibinfo {author} {\bibfnamefont
  {Z.}~\bibnamefont {Fang}}, \bibinfo {author} {\bibfnamefont {X.}~\bibnamefont
  {Dai}}, \bibinfo {author} {\bibfnamefont {Z.}~\bibnamefont {Hussain}}, \ and\
  \bibinfo {author} {\bibfnamefont {Y.~L.}\ \bibnamefont {Chen}},\ }\bibfield
  {title} {\enquote {\bibinfo {title} {Discovery of a three-dimensional
  topological dirac semimetal, {Na}$_3${Bi}},}\ }\href {\doibase
  10.1126/science.1245085} {\bibfield  {journal} {\bibinfo  {journal}
  {Science}\ }\textbf {\bibinfo {volume} {343}},\ \bibinfo {pages} {864--867}
  (\bibinfo {year} {2014}{\natexlab{a}})}\BibitemShut {NoStop}%
\bibitem [{\citenamefont {Wang}\ \emph {et~al.}(2013)\citenamefont {Wang},
  \citenamefont {Weng}, \citenamefont {Wu}, \citenamefont {Dai},\ and\
  \citenamefont {Fang}}]{Wang2013}%
  \BibitemOpen
  \bibfield  {author} {\bibinfo {author} {\bibfnamefont {Zhijun}\ \bibnamefont
  {Wang}}, \bibinfo {author} {\bibfnamefont {Hongming}\ \bibnamefont {Weng}},
  \bibinfo {author} {\bibfnamefont {Quansheng}\ \bibnamefont {Wu}}, \bibinfo
  {author} {\bibfnamefont {Xi}~\bibnamefont {Dai}}, \ and\ \bibinfo {author}
  {\bibfnamefont {Zhong}\ \bibnamefont {Fang}},\ }\bibfield  {title} {\enquote
  {\bibinfo {title} {Three-dimensional dirac semimetal and quantum transport in
  {Cd}$_3${As}$_2$},}\ }\href {\doibase 10.1103/physrevb.88.125427} {\bibfield
  {journal} {\bibinfo  {journal} {Phys. Rev. B}\ }\textbf {\bibinfo {volume}
  {88}},\ \bibinfo {pages} {125427} (\bibinfo {year} {2013})}\BibitemShut
  {NoStop}%
\bibitem [{\citenamefont {Liu}\ \emph {et~al.}(2014{\natexlab{b}})\citenamefont
  {Liu}, \citenamefont {Jiang}, \citenamefont {Zhou}, \citenamefont {Wang},
  \citenamefont {Zhang}, \citenamefont {Weng}, \citenamefont {Prabhakaran},
  \citenamefont {Mo}, \citenamefont {Peng}, \citenamefont {Dudin},
  \citenamefont {Kim}, \citenamefont {Hoesch}, \citenamefont {Fang},
  \citenamefont {Dai}, \citenamefont {Shen}, \citenamefont {Feng},
  \citenamefont {Hussain},\ and\ \citenamefont {Chen}}]{Liu2014}%
  \BibitemOpen
  \bibfield  {author} {\bibinfo {author} {\bibfnamefont {Z.~K.}\ \bibnamefont
  {Liu}}, \bibinfo {author} {\bibfnamefont {J.}~\bibnamefont {Jiang}}, \bibinfo
  {author} {\bibfnamefont {B.}~\bibnamefont {Zhou}}, \bibinfo {author}
  {\bibfnamefont {Z.~J.}\ \bibnamefont {Wang}}, \bibinfo {author}
  {\bibfnamefont {Y.}~\bibnamefont {Zhang}}, \bibinfo {author} {\bibfnamefont
  {H.~M.}\ \bibnamefont {Weng}}, \bibinfo {author} {\bibfnamefont
  {D.}~\bibnamefont {Prabhakaran}}, \bibinfo {author} {\bibfnamefont {S-K.}\
  \bibnamefont {Mo}}, \bibinfo {author} {\bibfnamefont {H.}~\bibnamefont
  {Peng}}, \bibinfo {author} {\bibfnamefont {P.}~\bibnamefont {Dudin}},
  \bibinfo {author} {\bibfnamefont {T.}~\bibnamefont {Kim}}, \bibinfo {author}
  {\bibfnamefont {M.}~\bibnamefont {Hoesch}}, \bibinfo {author} {\bibfnamefont
  {Z.}~\bibnamefont {Fang}}, \bibinfo {author} {\bibfnamefont {X.}~\bibnamefont
  {Dai}}, \bibinfo {author} {\bibfnamefont {Z.~X.}\ \bibnamefont {Shen}},
  \bibinfo {author} {\bibfnamefont {D.~L.}\ \bibnamefont {Feng}}, \bibinfo
  {author} {\bibfnamefont {Z.}~\bibnamefont {Hussain}}, \ and\ \bibinfo
  {author} {\bibfnamefont {Y.~L.}\ \bibnamefont {Chen}},\ }\bibfield  {title}
  {\enquote {\bibinfo {title} {A stable three-dimensional topological dirac
  semimetal {Cd}$_3${As}$_2$},}\ }\href {\doibase 10.1038/nmat3990} {\bibfield
  {journal} {\bibinfo  {journal} {Nat. Mater.}\ }\textbf {\bibinfo {volume}
  {13}},\ \bibinfo {pages} {677--681} (\bibinfo {year}
  {2014}{\natexlab{b}})}\BibitemShut {NoStop}%
\bibitem [{\citenamefont {Burkov}(2015)}]{Burkov2015}%
  \BibitemOpen
  \bibfield  {author} {\bibinfo {author} {\bibfnamefont {A~A}\ \bibnamefont
  {Burkov}},\ }\bibfield  {title} {\enquote {\bibinfo {title} {Chiral anomaly
  and transport in {W}eyl metals},}\ }\href {\doibase
  10.1088/0953-8984/27/11/113201} {\bibfield  {journal} {\bibinfo  {journal}
  {J. Phys.: Condens. Matter}\ }\textbf {\bibinfo {volume} {27}},\ \bibinfo
  {pages} {113201} (\bibinfo {year} {2015})}\BibitemShut {NoStop}%
\bibitem [{\citenamefont {Nielsen}\ and\ \citenamefont
  {Ninomiya}(1983)}]{Nielsen1983}%
  \BibitemOpen
  \bibfield  {author} {\bibinfo {author} {\bibfnamefont {H.B.}\ \bibnamefont
  {Nielsen}}\ and\ \bibinfo {author} {\bibfnamefont {Masao}\ \bibnamefont
  {Ninomiya}},\ }\bibfield  {title} {\enquote {\bibinfo {title} {The
  {A}dler-{B}ell-{J}ackiw anomaly and {W}eyl fermions in a crystal},}\ }\href
  {\doibase 10.1016/0370-2693(83)91529-0} {\bibfield  {journal} {\bibinfo
  {journal} {Phys. Lett. B}\ }\textbf {\bibinfo {volume} {130}},\ \bibinfo
  {pages} {389--396} (\bibinfo {year} {1983})}\BibitemShut {NoStop}%
\bibitem [{\citenamefont {Adler}(1969)}]{Adler1969}%
  \BibitemOpen
  \bibfield  {author} {\bibinfo {author} {\bibfnamefont {Stephen~L.}\
  \bibnamefont {Adler}},\ }\bibfield  {title} {\enquote {\bibinfo {title}
  {Axial-{V}ector {V}ertex in {S}pinor {E}lectrodynamics},}\ }\href {\doibase
  10.1103/physrev.177.2426} {\bibfield  {journal} {\bibinfo  {journal} {Phys.
  Rev.}\ }\textbf {\bibinfo {volume} {177}},\ \bibinfo {pages} {2426--2438}
  (\bibinfo {year} {1969})}\BibitemShut {NoStop}%
\bibitem [{\citenamefont {Bell}\ and\ \citenamefont {Jackiw}(1969)}]{Bell1969}%
  \BibitemOpen
  \bibfield  {author} {\bibinfo {author} {\bibfnamefont {J.~S.}\ \bibnamefont
  {Bell}}\ and\ \bibinfo {author} {\bibfnamefont {R.}~\bibnamefont {Jackiw}},\
  }\bibfield  {title} {\enquote {\bibinfo {title} {A {PCAC} puzzle: $\pi^0
  \rightarrow \gamma \gamma$ in the $\sigma$-model},}\ }\href {\doibase
  10.1007/bf02823296} {\bibfield  {journal} {\bibinfo  {journal} {Il Nuovo
  Cimento A}\ }\textbf {\bibinfo {volume} {60}},\ \bibinfo {pages} {47--61}
  (\bibinfo {year} {1969})}\BibitemShut {NoStop}%
\bibitem [{\citenamefont {Lv}\ \emph {et~al.}(2015)\citenamefont {Lv},
  \citenamefont {Weng}, \citenamefont {Fu}, \citenamefont {Wang}, \citenamefont
  {Miao}, \citenamefont {Ma}, \citenamefont {Richard}, \citenamefont {Huang},
  \citenamefont {Zhao}, \citenamefont {Chen}, \citenamefont {Fang},
  \citenamefont {Dai}, \citenamefont {Qian},\ and\ \citenamefont
  {Ding}}]{Lv2015}%
  \BibitemOpen
  \bibfield  {author} {\bibinfo {author} {\bibfnamefont
  {B.{\hspace{0.167em}}Q.}\ \bibnamefont {Lv}}, \bibinfo {author}
  {\bibfnamefont {H.{\hspace{0.167em}}M.}\ \bibnamefont {Weng}}, \bibinfo
  {author} {\bibfnamefont {B.{\hspace{0.167em}}B.}\ \bibnamefont {Fu}},
  \bibinfo {author} {\bibfnamefont {X.{\hspace{0.167em}}P.}\ \bibnamefont
  {Wang}}, \bibinfo {author} {\bibfnamefont {H.}~\bibnamefont {Miao}}, \bibinfo
  {author} {\bibfnamefont {J.}~\bibnamefont {Ma}}, \bibinfo {author}
  {\bibfnamefont {P.}~\bibnamefont {Richard}}, \bibinfo {author} {\bibfnamefont
  {X.{\hspace{0.167em}}C.}\ \bibnamefont {Huang}}, \bibinfo {author}
  {\bibfnamefont {L.{\hspace{0.167em}}X.}\ \bibnamefont {Zhao}}, \bibinfo
  {author} {\bibfnamefont {G.{\hspace{0.167em}}F.}\ \bibnamefont {Chen}},
  \bibinfo {author} {\bibfnamefont {Z.}~\bibnamefont {Fang}}, \bibinfo {author}
  {\bibfnamefont {X.}~\bibnamefont {Dai}}, \bibinfo {author} {\bibfnamefont
  {T.}~\bibnamefont {Qian}}, \ and\ \bibinfo {author} {\bibfnamefont
  {H.}~\bibnamefont {Ding}},\ }\bibfield  {title} {\enquote {\bibinfo {title}
  {{Experimental Discovery of Weyl Semimetal {TaAs}}},}\ }\href {\doibase
  10.1103/physrevx.5.031013} {\bibfield  {journal} {\bibinfo  {journal} {Phys.
  Rev. X}\ }\textbf {\bibinfo {volume} {5}},\ \bibinfo {pages} {031013}
  (\bibinfo {year} {2015})}\BibitemShut {NoStop}%
\bibitem [{\citenamefont {Huang}\ \emph
  {et~al.}(2015{\natexlab{a}})\citenamefont {Huang}, \citenamefont {Xu},
  \citenamefont {Belopolski}, \citenamefont {Lee}, \citenamefont {Chang},
  \citenamefont {Wang}, \citenamefont {Alidoust}, \citenamefont {Bian},
  \citenamefont {Neupane}, \citenamefont {Zhang}, \citenamefont {Jia},
  \citenamefont {Bansil}, \citenamefont {Lin},\ and\ \citenamefont
  {Hasan}}]{Huang2015}%
  \BibitemOpen
  \bibfield  {author} {\bibinfo {author} {\bibfnamefont {Shin-Ming}\
  \bibnamefont {Huang}}, \bibinfo {author} {\bibfnamefont {Su-Yang}\
  \bibnamefont {Xu}}, \bibinfo {author} {\bibfnamefont {Ilya}\ \bibnamefont
  {Belopolski}}, \bibinfo {author} {\bibfnamefont {Chi-Cheng}\ \bibnamefont
  {Lee}}, \bibinfo {author} {\bibfnamefont {Guoqing}\ \bibnamefont {Chang}},
  \bibinfo {author} {\bibfnamefont {BaoKai}\ \bibnamefont {Wang}}, \bibinfo
  {author} {\bibfnamefont {Nasser}\ \bibnamefont {Alidoust}}, \bibinfo {author}
  {\bibfnamefont {Guang}\ \bibnamefont {Bian}}, \bibinfo {author}
  {\bibfnamefont {Madhab}\ \bibnamefont {Neupane}}, \bibinfo {author}
  {\bibfnamefont {Chenglong}\ \bibnamefont {Zhang}}, \bibinfo {author}
  {\bibfnamefont {Shuang}\ \bibnamefont {Jia}}, \bibinfo {author}
  {\bibfnamefont {Arun}\ \bibnamefont {Bansil}}, \bibinfo {author}
  {\bibfnamefont {Hsin}\ \bibnamefont {Lin}}, \ and\ \bibinfo {author}
  {\bibfnamefont {M.~Zahid}\ \bibnamefont {Hasan}},\ }\bibfield  {title}
  {\enquote {\bibinfo {title} {{A Weyl Fermion semimetal with surface Fermi
  arcs in the transition metal monopnictide {TaAs} class}},}\ }\href {\doibase
  10.1038/ncomms8373} {\bibfield  {journal} {\bibinfo  {journal} {Nat.
  Commun.}\ }\textbf {\bibinfo {volume} {6}},\ \bibinfo {pages} {7373}
  (\bibinfo {year} {2015}{\natexlab{a}})}\BibitemShut {NoStop}%
\bibitem [{\citenamefont {Deng}\ \emph {et~al.}(2016)\citenamefont {Deng},
  \citenamefont {Wan}, \citenamefont {Deng}, \citenamefont {Zhang},
  \citenamefont {Ding}, \citenamefont {Wang}, \citenamefont {Yan},
  \citenamefont {Huang}, \citenamefont {Zhang}, \citenamefont {Xu},
  \citenamefont {Denlinger}, \citenamefont {Fedorov}, \citenamefont {Yang},
  \citenamefont {Duan}, \citenamefont {Yao}, \citenamefont {Wu}, \citenamefont
  {Fan}, \citenamefont {Zhang}, \citenamefont {Chen},\ and\ \citenamefont
  {Zhou}}]{Deng2016}%
  \BibitemOpen
  \bibfield  {author} {\bibinfo {author} {\bibfnamefont {Ke}~\bibnamefont
  {Deng}}, \bibinfo {author} {\bibfnamefont {Guoliang}\ \bibnamefont {Wan}},
  \bibinfo {author} {\bibfnamefont {Peng}\ \bibnamefont {Deng}}, \bibinfo
  {author} {\bibfnamefont {Kenan}\ \bibnamefont {Zhang}}, \bibinfo {author}
  {\bibfnamefont {Shijie}\ \bibnamefont {Ding}}, \bibinfo {author}
  {\bibfnamefont {Eryin}\ \bibnamefont {Wang}}, \bibinfo {author}
  {\bibfnamefont {Mingzhe}\ \bibnamefont {Yan}}, \bibinfo {author}
  {\bibfnamefont {Huaqing}\ \bibnamefont {Huang}}, \bibinfo {author}
  {\bibfnamefont {Hongyun}\ \bibnamefont {Zhang}}, \bibinfo {author}
  {\bibfnamefont {Zhilin}\ \bibnamefont {Xu}}, \bibinfo {author} {\bibfnamefont
  {Jonathan}\ \bibnamefont {Denlinger}}, \bibinfo {author} {\bibfnamefont
  {Alexei}\ \bibnamefont {Fedorov}}, \bibinfo {author} {\bibfnamefont {Haitao}\
  \bibnamefont {Yang}}, \bibinfo {author} {\bibfnamefont {Wenhui}\ \bibnamefont
  {Duan}}, \bibinfo {author} {\bibfnamefont {Hong}\ \bibnamefont {Yao}},
  \bibinfo {author} {\bibfnamefont {Yang}\ \bibnamefont {Wu}}, \bibinfo
  {author} {\bibfnamefont {Shoushan}\ \bibnamefont {Fan}}, \bibinfo {author}
  {\bibfnamefont {Haijun}\ \bibnamefont {Zhang}}, \bibinfo {author}
  {\bibfnamefont {Xi}~\bibnamefont {Chen}}, \ and\ \bibinfo {author}
  {\bibfnamefont {Shuyun}\ \bibnamefont {Zhou}},\ }\bibfield  {title} {\enquote
  {\bibinfo {title} {{Experimental observation of topological Fermi arcs in
  type-{II} Weyl semimetal {MoTe}$_2$}},}\ }\href {\doibase 10.1038/nphys3871}
  {\bibfield  {journal} {\bibinfo  {journal} {Nat. Phys.}\ }\textbf {\bibinfo
  {volume} {12}},\ \bibinfo {pages} {1105--1110} (\bibinfo {year}
  {2016})}\BibitemShut {NoStop}%
\bibitem [{\citenamefont {Xu}\ \emph {et~al.}(2016)\citenamefont {Xu},
  \citenamefont {Weng}, \citenamefont {Lv}, \citenamefont {Matt}, \citenamefont
  {Park}, \citenamefont {Bisti}, \citenamefont {Strocov}, \citenamefont
  {Gawryluk}, \citenamefont {Pomjakushina}, \citenamefont {Conder},
  \citenamefont {Plumb}, \citenamefont {Radovic}, \citenamefont {Aut{\`{e}}s},
  \citenamefont {Yazyev}, \citenamefont {Fang}, \citenamefont {Dai},
  \citenamefont {Qian}, \citenamefont {Mesot}, \citenamefont {Ding},\ and\
  \citenamefont {Shi}}]{Xu2016}%
  \BibitemOpen
  \bibfield  {author} {\bibinfo {author} {\bibfnamefont {N.}~\bibnamefont
  {Xu}}, \bibinfo {author} {\bibfnamefont {H.~M.}\ \bibnamefont {Weng}},
  \bibinfo {author} {\bibfnamefont {B.~Q.}\ \bibnamefont {Lv}}, \bibinfo
  {author} {\bibfnamefont {C.~E.}\ \bibnamefont {Matt}}, \bibinfo {author}
  {\bibfnamefont {J.}~\bibnamefont {Park}}, \bibinfo {author} {\bibfnamefont
  {F.}~\bibnamefont {Bisti}}, \bibinfo {author} {\bibfnamefont {V.~N.}\
  \bibnamefont {Strocov}}, \bibinfo {author} {\bibfnamefont {D.}~\bibnamefont
  {Gawryluk}}, \bibinfo {author} {\bibfnamefont {E.}~\bibnamefont
  {Pomjakushina}}, \bibinfo {author} {\bibfnamefont {K.}~\bibnamefont
  {Conder}}, \bibinfo {author} {\bibfnamefont {N.~C.}\ \bibnamefont {Plumb}},
  \bibinfo {author} {\bibfnamefont {M.}~\bibnamefont {Radovic}}, \bibinfo
  {author} {\bibfnamefont {G.}~\bibnamefont {Aut{\`{e}}s}}, \bibinfo {author}
  {\bibfnamefont {O.~V.}\ \bibnamefont {Yazyev}}, \bibinfo {author}
  {\bibfnamefont {Z.}~\bibnamefont {Fang}}, \bibinfo {author} {\bibfnamefont
  {X.}~\bibnamefont {Dai}}, \bibinfo {author} {\bibfnamefont {T.}~\bibnamefont
  {Qian}}, \bibinfo {author} {\bibfnamefont {J.}~\bibnamefont {Mesot}},
  \bibinfo {author} {\bibfnamefont {H.}~\bibnamefont {Ding}}, \ and\ \bibinfo
  {author} {\bibfnamefont {M.}~\bibnamefont {Shi}},\ }\bibfield  {title}
  {\enquote {\bibinfo {title} {{Observation of Weyl nodes and Fermi arcs in
  tantalum phosphide}},}\ }\href {\doibase 10.1038/ncomms11006} {\bibfield
  {journal} {\bibinfo  {journal} {Nat. Commun.}\ }\textbf {\bibinfo {volume}
  {7}},\ \bibinfo {pages} {11006} (\bibinfo {year} {2016})}\BibitemShut
  {NoStop}%
\bibitem [{\citenamefont {Huang}\ \emph
  {et~al.}(2015{\natexlab{b}})\citenamefont {Huang}, \citenamefont {Zhao},
  \citenamefont {Long}, \citenamefont {Wang}, \citenamefont {Chen},
  \citenamefont {Yang}, \citenamefont {Liang}, \citenamefont {Xue},
  \citenamefont {Weng}, \citenamefont {Fang}, \citenamefont {Dai},\ and\
  \citenamefont {Chen}}]{Huang2015a}%
  \BibitemOpen
  \bibfield  {author} {\bibinfo {author} {\bibfnamefont {Xiaochun}\
  \bibnamefont {Huang}}, \bibinfo {author} {\bibfnamefont {Lingxiao}\
  \bibnamefont {Zhao}}, \bibinfo {author} {\bibfnamefont {Yujia}\ \bibnamefont
  {Long}}, \bibinfo {author} {\bibfnamefont {Peipei}\ \bibnamefont {Wang}},
  \bibinfo {author} {\bibfnamefont {Dong}\ \bibnamefont {Chen}}, \bibinfo
  {author} {\bibfnamefont {Zhanhai}\ \bibnamefont {Yang}}, \bibinfo {author}
  {\bibfnamefont {Hui}\ \bibnamefont {Liang}}, \bibinfo {author} {\bibfnamefont
  {Mianqi}\ \bibnamefont {Xue}}, \bibinfo {author} {\bibfnamefont {Hongming}\
  \bibnamefont {Weng}}, \bibinfo {author} {\bibfnamefont {Zhong}\ \bibnamefont
  {Fang}}, \bibinfo {author} {\bibfnamefont {Xi}~\bibnamefont {Dai}}, \ and\
  \bibinfo {author} {\bibfnamefont {Genfu}\ \bibnamefont {Chen}},\ }\bibfield
  {title} {\enquote {\bibinfo {title} {{Observation of the
  Chiral-Anomaly-Induced Negative Magnetoresistance in 3D Weyl Semimetal
  {TaAs}}},}\ }\href {\doibase 10.1103/physrevx.5.031023} {\bibfield  {journal}
  {\bibinfo  {journal} {Phys. Rev. X}\ }\textbf {\bibinfo {volume} {5}},\
  \bibinfo {pages} {031023} (\bibinfo {year} {2015}{\natexlab{b}})}\BibitemShut
  {NoStop}%
\bibitem [{\citenamefont {Xiong}\ \emph {et~al.}(2015)\citenamefont {Xiong},
  \citenamefont {Kushwaha}, \citenamefont {Liang}, \citenamefont {Krizan},
  \citenamefont {Hirschberger}, \citenamefont {Wang}, \citenamefont {Cava},\
  and\ \citenamefont {Ong}}]{Xiong2015}%
  \BibitemOpen
  \bibfield  {author} {\bibinfo {author} {\bibfnamefont {J.}~\bibnamefont
  {Xiong}}, \bibinfo {author} {\bibfnamefont {S.~K.}\ \bibnamefont {Kushwaha}},
  \bibinfo {author} {\bibfnamefont {T.}~\bibnamefont {Liang}}, \bibinfo
  {author} {\bibfnamefont {J.~W.}\ \bibnamefont {Krizan}}, \bibinfo {author}
  {\bibfnamefont {M.}~\bibnamefont {Hirschberger}}, \bibinfo {author}
  {\bibfnamefont {W.}~\bibnamefont {Wang}}, \bibinfo {author} {\bibfnamefont
  {R.~J.}\ \bibnamefont {Cava}}, \ and\ \bibinfo {author} {\bibfnamefont
  {N.~P.}\ \bibnamefont {Ong}},\ }\bibfield  {title} {\enquote {\bibinfo
  {title} {{Evidence for the chiral anomaly in the Dirac semimetal
  Na$_3$Bi}},}\ }\href {\doibase 10.1126/science.aac6089} {\bibfield  {journal}
  {\bibinfo  {journal} {Science}\ }\textbf {\bibinfo {volume} {350}},\ \bibinfo
  {pages} {413--416} (\bibinfo {year} {2015})}\BibitemShut {NoStop}%
\bibitem [{\citenamefont {Li}\ \emph {et~al.}(2015)\citenamefont {Li},
  \citenamefont {Wang}, \citenamefont {Liu}, \citenamefont {Wang},
  \citenamefont {Liao},\ and\ \citenamefont {Yu}}]{Li2015}%
  \BibitemOpen
  \bibfield  {author} {\bibinfo {author} {\bibfnamefont {Cai-Zhen}\
  \bibnamefont {Li}}, \bibinfo {author} {\bibfnamefont {Li-Xian}\ \bibnamefont
  {Wang}}, \bibinfo {author} {\bibfnamefont {Haiwen}\ \bibnamefont {Liu}},
  \bibinfo {author} {\bibfnamefont {Jian}\ \bibnamefont {Wang}}, \bibinfo
  {author} {\bibfnamefont {Zhi-Min}\ \bibnamefont {Liao}}, \ and\ \bibinfo
  {author} {\bibfnamefont {Da-Peng}\ \bibnamefont {Yu}},\ }\bibfield  {title}
  {\enquote {\bibinfo {title} {{Giant negative magnetoresistance induced by the
  chiral anomaly in individual Cd$_3$As$_2$ nanowires}},}\ }\href {\doibase
  10.1038/ncomms10137} {\bibfield  {journal} {\bibinfo  {journal} {Nat.
  Commun.}\ }\textbf {\bibinfo {volume} {6}},\ \bibinfo {pages} {10137}
  (\bibinfo {year} {2015})}\BibitemShut {NoStop}%
\bibitem [{\citenamefont {Zhang}\ \emph {et~al.}(2016)\citenamefont {Zhang},
  \citenamefont {Xu}, \citenamefont {Belopolski}, \citenamefont {Yuan},
  \citenamefont {Lin}, \citenamefont {Tong}, \citenamefont {Bian},
  \citenamefont {Alidoust}, \citenamefont {Lee}, \citenamefont {Huang},
  \citenamefont {Chang}, \citenamefont {Chang}, \citenamefont {Hsu},
  \citenamefont {Jeng}, \citenamefont {Neupane}, \citenamefont {Sanchez},
  \citenamefont {Zheng}, \citenamefont {Wang}, \citenamefont {Lin},
  \citenamefont {Zhang}, \citenamefont {Lu}, \citenamefont {Shen},
  \citenamefont {Neupert}, \citenamefont {Hasan},\ and\ \citenamefont
  {Jia}}]{Zhang2016}%
  \BibitemOpen
  \bibfield  {author} {\bibinfo {author} {\bibfnamefont {Cheng-Long}\
  \bibnamefont {Zhang}}, \bibinfo {author} {\bibfnamefont {Su-Yang}\
  \bibnamefont {Xu}}, \bibinfo {author} {\bibfnamefont {Ilya}\ \bibnamefont
  {Belopolski}}, \bibinfo {author} {\bibfnamefont {Zhujun}\ \bibnamefont
  {Yuan}}, \bibinfo {author} {\bibfnamefont {Ziquan}\ \bibnamefont {Lin}},
  \bibinfo {author} {\bibfnamefont {Bingbing}\ \bibnamefont {Tong}}, \bibinfo
  {author} {\bibfnamefont {Guang}\ \bibnamefont {Bian}}, \bibinfo {author}
  {\bibfnamefont {Nasser}\ \bibnamefont {Alidoust}}, \bibinfo {author}
  {\bibfnamefont {Chi-Cheng}\ \bibnamefont {Lee}}, \bibinfo {author}
  {\bibfnamefont {Shin-Ming}\ \bibnamefont {Huang}}, \bibinfo {author}
  {\bibfnamefont {Tay-Rong}\ \bibnamefont {Chang}}, \bibinfo {author}
  {\bibfnamefont {Guoqing}\ \bibnamefont {Chang}}, \bibinfo {author}
  {\bibfnamefont {Chuang-Han}\ \bibnamefont {Hsu}}, \bibinfo {author}
  {\bibfnamefont {Horng-Tay}\ \bibnamefont {Jeng}}, \bibinfo {author}
  {\bibfnamefont {Madhab}\ \bibnamefont {Neupane}}, \bibinfo {author}
  {\bibfnamefont {Daniel~S.}\ \bibnamefont {Sanchez}}, \bibinfo {author}
  {\bibfnamefont {Hao}\ \bibnamefont {Zheng}}, \bibinfo {author} {\bibfnamefont
  {Junfeng}\ \bibnamefont {Wang}}, \bibinfo {author} {\bibfnamefont {Hsin}\
  \bibnamefont {Lin}}, \bibinfo {author} {\bibfnamefont {Chi}\ \bibnamefont
  {Zhang}}, \bibinfo {author} {\bibfnamefont {Hai-Zhou}\ \bibnamefont {Lu}},
  \bibinfo {author} {\bibfnamefont {Shun-Qing}\ \bibnamefont {Shen}}, \bibinfo
  {author} {\bibfnamefont {Titus}\ \bibnamefont {Neupert}}, \bibinfo {author}
  {\bibfnamefont {M.~Zahid}\ \bibnamefont {Hasan}}, \ and\ \bibinfo {author}
  {\bibfnamefont {Shuang}\ \bibnamefont {Jia}},\ }\bibfield  {title} {\enquote
  {\bibinfo {title} {{Signatures of the
  Adler{\textendash}Bell{\textendash}Jackiw chiral anomaly in a Weyl fermion
  semimetal}},}\ }\href {\doibase 10.1038/ncomms10735} {\bibfield  {journal}
  {\bibinfo  {journal} {Nat. Commun.}\ }\textbf {\bibinfo {volume} {7}},\
  \bibinfo {pages} {10735} (\bibinfo {year} {2016})}\BibitemShut {NoStop}%
\bibitem [{\citenamefont {Hirschberger}\ \emph {et~al.}(2016)\citenamefont
  {Hirschberger}, \citenamefont {Kushwaha}, \citenamefont {Wang}, \citenamefont
  {Gibson}, \citenamefont {Liang}, \citenamefont {Belvin}, \citenamefont
  {Bernevig}, \citenamefont {Cava},\ and\ \citenamefont
  {Ong}}]{Hirschberger2016}%
  \BibitemOpen
  \bibfield  {author} {\bibinfo {author} {\bibfnamefont {Max}\ \bibnamefont
  {Hirschberger}}, \bibinfo {author} {\bibfnamefont {Satya}\ \bibnamefont
  {Kushwaha}}, \bibinfo {author} {\bibfnamefont {Zhijun}\ \bibnamefont {Wang}},
  \bibinfo {author} {\bibfnamefont {Quinn}\ \bibnamefont {Gibson}}, \bibinfo
  {author} {\bibfnamefont {Sihang}\ \bibnamefont {Liang}}, \bibinfo {author}
  {\bibfnamefont {Carina~A.}\ \bibnamefont {Belvin}}, \bibinfo {author}
  {\bibfnamefont {B.~A.}\ \bibnamefont {Bernevig}}, \bibinfo {author}
  {\bibfnamefont {R.~J.}\ \bibnamefont {Cava}}, \ and\ \bibinfo {author}
  {\bibfnamefont {N.~P.}\ \bibnamefont {Ong}},\ }\bibfield  {title} {\enquote
  {\bibinfo {title} {{The chiral anomaly and thermopower of Weyl fermions in
  the half-Heusler~{GdPtBi}}},}\ }\href {\doibase 10.1038/nmat4684} {\bibfield
  {journal} {\bibinfo  {journal} {Nat. Mater.}\ }\textbf {\bibinfo {volume}
  {15}},\ \bibinfo {pages} {1161} (\bibinfo {year} {2016})}\BibitemShut
  {NoStop}%
\bibitem [{\citenamefont {Young}\ \emph {et~al.}(2012)\citenamefont {Young},
  \citenamefont {Zaheer}, \citenamefont {Teo}, \citenamefont {Kane},
  \citenamefont {Mele},\ and\ \citenamefont {Rappe}}]{Young2012a}%
  \BibitemOpen
  \bibfield  {author} {\bibinfo {author} {\bibfnamefont {S.~M.}\ \bibnamefont
  {Young}}, \bibinfo {author} {\bibfnamefont {S.}~\bibnamefont {Zaheer}},
  \bibinfo {author} {\bibfnamefont {J.~C.~Y.}\ \bibnamefont {Teo}}, \bibinfo
  {author} {\bibfnamefont {C.~L.}\ \bibnamefont {Kane}}, \bibinfo {author}
  {\bibfnamefont {E.~J.}\ \bibnamefont {Mele}}, \ and\ \bibinfo {author}
  {\bibfnamefont {A.~M.}\ \bibnamefont {Rappe}},\ }\bibfield  {title} {\enquote
  {\bibinfo {title} {{Dirac Semimetal in Three Dimensions}},}\ }\href {\doibase
  10.1103/physrevlett.108.140405} {\bibfield  {journal} {\bibinfo  {journal}
  {Phys. Rev. Lett.}\ }\textbf {\bibinfo {volume} {108}},\ \bibinfo {pages}
  {140405} (\bibinfo {year} {2012})}\BibitemShut {NoStop}%
\bibitem [{\citenamefont {Ma{\~{n}}es}(2012)}]{Manes2012}%
  \BibitemOpen
  \bibfield  {author} {\bibinfo {author} {\bibfnamefont {J.~L.}\ \bibnamefont
  {Ma{\~{n}}es}},\ }\bibfield  {title} {\enquote {\bibinfo {title} {Existence
  of bulk chiral fermions and crystal symmetry},}\ }\href {\doibase
  10.1103/physrevb.85.155118} {\bibfield  {journal} {\bibinfo  {journal} {Phys.
  Rev. B}\ }\textbf {\bibinfo {volume} {85}},\ \bibinfo {pages} {155118}
  (\bibinfo {year} {2012})}\BibitemShut {NoStop}%
\bibitem [{\citenamefont {Yang}\ and\ \citenamefont
  {Nagaosa}(2014)}]{Yang2014}%
  \BibitemOpen
  \bibfield  {author} {\bibinfo {author} {\bibfnamefont {Bohm-Jung}\
  \bibnamefont {Yang}}\ and\ \bibinfo {author} {\bibfnamefont {Naoto}\
  \bibnamefont {Nagaosa}},\ }\bibfield  {title} {\enquote {\bibinfo {title}
  {{Classification of stable three-dimensional Dirac semimetals with nontrivial
  topology}},}\ }\href {\doibase 10.1038/ncomms5898} {\bibfield  {journal}
  {\bibinfo  {journal} {Nat. Commun.}\ }\textbf {\bibinfo {volume} {5}},\
  \bibinfo {pages} {4898} (\bibinfo {year} {2014})}\BibitemShut {NoStop}%
\bibitem [{\citenamefont {Song}\ \emph {et~al.}(2018)\citenamefont {Song},
  \citenamefont {Zhang},\ and\ \citenamefont {Fang}}]{Song2018}%
  \BibitemOpen
  \bibfield  {author} {\bibinfo {author} {\bibfnamefont {Zhida}\ \bibnamefont
  {Song}}, \bibinfo {author} {\bibfnamefont {Tiantian}\ \bibnamefont {Zhang}},
  \ and\ \bibinfo {author} {\bibfnamefont {Chen}\ \bibnamefont {Fang}},\
  }\bibfield  {title} {\enquote {\bibinfo {title} {{Diagnosis for Nonmagnetic
  Topological Semimetals in the Absence of Spin-Orbital Coupling}},}\ }\href
  {\doibase 10.1103/physrevx.8.031069} {\bibfield  {journal} {\bibinfo
  {journal} {Phys. Rev. X}\ }\textbf {\bibinfo {volume} {8}},\ \bibinfo {pages}
  {031069} (\bibinfo {year} {2018})}\BibitemShut {NoStop}%
\bibitem [{\citenamefont {Soluyanov}\ \emph {et~al.}(2015)\citenamefont
  {Soluyanov}, \citenamefont {Gresch}, \citenamefont {Wang}, \citenamefont
  {Wu}, \citenamefont {Troyer}, \citenamefont {Dai},\ and\ \citenamefont
  {Bernevig}}]{Soluyanov2015}%
  \BibitemOpen
  \bibfield  {author} {\bibinfo {author} {\bibfnamefont {Alexey~A.}\
  \bibnamefont {Soluyanov}}, \bibinfo {author} {\bibfnamefont {Dominik}\
  \bibnamefont {Gresch}}, \bibinfo {author} {\bibfnamefont {Zhijun}\
  \bibnamefont {Wang}}, \bibinfo {author} {\bibfnamefont {QuanSheng}\
  \bibnamefont {Wu}}, \bibinfo {author} {\bibfnamefont {Matthias}\ \bibnamefont
  {Troyer}}, \bibinfo {author} {\bibfnamefont {Xi}~\bibnamefont {Dai}}, \ and\
  \bibinfo {author} {\bibfnamefont {B.~Andrei}\ \bibnamefont {Bernevig}},\
  }\bibfield  {title} {\enquote {\bibinfo {title} {{Type-{II} Weyl
  semimetals}},}\ }\href {\doibase 10.1038/nature15768} {\bibfield  {journal}
  {\bibinfo  {journal} {Nature}\ }\textbf {\bibinfo {volume} {527}},\ \bibinfo
  {pages} {495--498} (\bibinfo {year} {2015})}\BibitemShut {NoStop}%
\bibitem [{\citenamefont {Kajita}\ \emph {et~al.}(2014)\citenamefont {Kajita},
  \citenamefont {Nishio}, \citenamefont {Tajima}, \citenamefont {Suzumura},\
  and\ \citenamefont {Kobayashi}}]{Kajita2014}%
  \BibitemOpen
  \bibfield  {author} {\bibinfo {author} {\bibfnamefont {Koji}\ \bibnamefont
  {Kajita}}, \bibinfo {author} {\bibfnamefont {Yutaka}\ \bibnamefont {Nishio}},
  \bibinfo {author} {\bibfnamefont {Naoya}\ \bibnamefont {Tajima}}, \bibinfo
  {author} {\bibfnamefont {Yoshikazu}\ \bibnamefont {Suzumura}}, \ and\
  \bibinfo {author} {\bibfnamefont {Akito}\ \bibnamefont {Kobayashi}},\
  }\bibfield  {title} {\enquote {\bibinfo {title} {{Molecular Dirac Fermion
  Systems {\textemdash} Theoretical and Experimental Approaches
  {\textemdash}}},}\ }\href {\doibase 10.7566/jpsj.83.072002} {\bibfield
  {journal} {\bibinfo  {journal} {J. Phys. Soc. Jpn.}\ }\textbf {\bibinfo
  {volume} {83}},\ \bibinfo {pages} {072002} (\bibinfo {year}
  {2014})}\BibitemShut {NoStop}%
\bibitem [{\citenamefont {Zhang}\ \emph {et~al.}(2019)\citenamefont {Zhang},
  \citenamefont {Jiang}, \citenamefont {Song}, \citenamefont {Huang},
  \citenamefont {He}, \citenamefont {Fang}, \citenamefont {Weng},\ and\
  \citenamefont {Fang}}]{Zhang2019}%
  \BibitemOpen
  \bibfield  {author} {\bibinfo {author} {\bibfnamefont {Tiantian}\
  \bibnamefont {Zhang}}, \bibinfo {author} {\bibfnamefont {Yi}~\bibnamefont
  {Jiang}}, \bibinfo {author} {\bibfnamefont {Zhida}\ \bibnamefont {Song}},
  \bibinfo {author} {\bibfnamefont {He}~\bibnamefont {Huang}}, \bibinfo
  {author} {\bibfnamefont {Yuqing}\ \bibnamefont {He}}, \bibinfo {author}
  {\bibfnamefont {Zhong}\ \bibnamefont {Fang}}, \bibinfo {author}
  {\bibfnamefont {Hongming}\ \bibnamefont {Weng}}, \ and\ \bibinfo {author}
  {\bibfnamefont {Chen}\ \bibnamefont {Fang}},\ }\bibfield  {title} {\enquote
  {\bibinfo {title} {Catalogue of topological electronic materials},}\ }\href
  {\doibase 10.1038/s41586-019-0944-6} {\bibfield  {journal} {\bibinfo
  {journal} {Nature}\ }\textbf {\bibinfo {volume} {566}},\ \bibinfo {pages}
  {475--479} (\bibinfo {year} {2019})}\BibitemShut {NoStop}%
\bibitem [{\citenamefont {Bender}\ \emph {et~al.}(1984)\citenamefont {Bender},
  \citenamefont {Hennig}, \citenamefont {Schweitzer}, \citenamefont {Dietz},
  \citenamefont {Endres},\ and\ \citenamefont {Keller}}]{Bender1984}%
  \BibitemOpen
  \bibfield  {author} {\bibinfo {author} {\bibfnamefont {K.}~\bibnamefont
  {Bender}}, \bibinfo {author} {\bibfnamefont {I.}~\bibnamefont {Hennig}},
  \bibinfo {author} {\bibfnamefont {D.}~\bibnamefont {Schweitzer}}, \bibinfo
  {author} {\bibfnamefont {K.}~\bibnamefont {Dietz}}, \bibinfo {author}
  {\bibfnamefont {H.}~\bibnamefont {Endres}}, \ and\ \bibinfo {author}
  {\bibfnamefont {H.~J.}\ \bibnamefont {Keller}},\ }\bibfield  {title}
  {\enquote {\bibinfo {title} {{Synthesis, Structure and Physical Properties of
  a Two-Dimensional Organic Metal, Di[bis(ethylenedithiolo)tetrathiofulvalene]
  triiodide, (BEDT-TTF)$^+_2$I$^-_3$}},}\ }\href@noop {} {\bibfield  {journal}
  {\bibinfo  {journal} {Mol. Cryst. Liq. Cryst.}\ }\textbf {\bibinfo {volume}
  {108}},\ \bibinfo {pages} {359} (\bibinfo {year} {1984})}\BibitemShut
  {NoStop}%
\bibitem [{\citenamefont {Mori}\ \emph {et~al.}(1984)\citenamefont {Mori},
  \citenamefont {Kobayashi}, \citenamefont {Sasaki}, \citenamefont {Kobayashi},
  \citenamefont {Saito},\ and\ \citenamefont {Inokuchi}}]{Mori1984}%
  \BibitemOpen
  \bibfield  {author} {\bibinfo {author} {\bibfnamefont {T.}~\bibnamefont
  {Mori}}, \bibinfo {author} {\bibfnamefont {A.}~\bibnamefont {Kobayashi}},
  \bibinfo {author} {\bibfnamefont {Y.}~\bibnamefont {Sasaki}}, \bibinfo
  {author} {\bibfnamefont {H.}~\bibnamefont {Kobayashi}}, \bibinfo {author}
  {\bibfnamefont {G.}~\bibnamefont {Saito}}, \ and\ \bibinfo {author}
  {\bibfnamefont {H.}~\bibnamefont {Inokuchi}},\ }\bibfield  {title} {\enquote
  {\bibinfo {title} {{Band structures of two types of (BEDT-TTF)$_2$I$_3$}},}\
  }\href@noop {} {\bibfield  {journal} {\bibinfo  {journal} {Chem. Lett.}\
  }\textbf {\bibinfo {volume} {13}},\ \bibinfo {pages} {957--960} (\bibinfo
  {year} {1984})}\BibitemShut {NoStop}%
\bibitem [{\citenamefont {Kartsovnik}\ \emph {et~al.}(1985)\citenamefont
  {Kartsovnik}, \citenamefont {Kononovich}, \citenamefont {Laukin},
  \citenamefont {Khomenko},\ and\ \citenamefont {Schegolev}}]{Kartsovnik85}%
  \BibitemOpen
  \bibfield  {author} {\bibinfo {author} {\bibfnamefont {M.~V.}\ \bibnamefont
  {Kartsovnik}}, \bibinfo {author} {\bibfnamefont {P.~A.}\ \bibnamefont
  {Kononovich}}, \bibinfo {author} {\bibfnamefont {V.~N.}\ \bibnamefont
  {Laukin}}, \bibinfo {author} {\bibfnamefont {A.~G.}\ \bibnamefont
  {Khomenko}}, \ and\ \bibinfo {author} {\bibfnamefont {I.~F.}\ \bibnamefont
  {Schegolev}},\ }\bibfield  {title} {\enquote {\bibinfo {title}
  {{Investigation of the $T$-$P$ phase diagram for
  $\alpha$-(BEDT-TTF)$_2$I$_3$}},}\ }\href@noop {} {\bibfield  {journal}
  {\bibinfo  {journal} {Sov. Phys. JETP}\ }\textbf {\bibinfo {volume} {61}},\
  \bibinfo {pages} {866} (\bibinfo {year} {1985})}\BibitemShut {NoStop}%
\bibitem [{\citenamefont {Schwenk}\ \emph {et~al.}(1985)\citenamefont
  {Schwenk}, \citenamefont {Gross}, \citenamefont {Heidmann}, \citenamefont
  {Andres}, \citenamefont {Schweitzer},\ and\ \citenamefont
  {Keller}}]{Schwenk1985}%
  \BibitemOpen
  \bibfield  {author} {\bibinfo {author} {\bibfnamefont {Helmut}\ \bibnamefont
  {Schwenk}}, \bibinfo {author} {\bibfnamefont {Frieder}\ \bibnamefont
  {Gross}}, \bibinfo {author} {\bibfnamefont {Claus-Peter}\ \bibnamefont
  {Heidmann}}, \bibinfo {author} {\bibfnamefont {Klaus}\ \bibnamefont
  {Andres}}, \bibinfo {author} {\bibfnamefont {Dieter}\ \bibnamefont
  {Schweitzer}}, \ and\ \bibinfo {author} {\bibfnamefont {Heimo}\ \bibnamefont
  {Keller}},\ }\bibfield  {title} {\enquote {\bibinfo {title} {{$\alpha$- And
  $\beta$-(Bedt-{TTF})$_2$I$_3$ {\textendash} Two Modifications With
  Contrasting Groundstate Properties: Insulator and Volume Superconductor}},}\
  }\href {\doibase 10.1080/00268948508075181} {\bibfield  {journal} {\bibinfo
  {journal} {Mol. Cryst. Liq. Cryst.}\ }\textbf {\bibinfo {volume} {119}},\
  \bibinfo {pages} {329} (\bibinfo {year} {1985})}\BibitemShut {NoStop}%
\bibitem [{\citenamefont {Kajita}\ \emph {et~al.}(1992)\citenamefont {Kajita},
  \citenamefont {Ojiro}, \citenamefont {Fujii}, \citenamefont {Nishio},
  \citenamefont {Kobayashi}, \citenamefont {Kobayashi},\ and\ \citenamefont
  {Kato}}]{Kajita1992}%
  \BibitemOpen
  \bibfield  {author} {\bibinfo {author} {\bibfnamefont {Koji}\ \bibnamefont
  {Kajita}}, \bibinfo {author} {\bibfnamefont {Tukasa}\ \bibnamefont {Ojiro}},
  \bibinfo {author} {\bibfnamefont {Hideharu}\ \bibnamefont {Fujii}}, \bibinfo
  {author} {\bibfnamefont {Yutaka}\ \bibnamefont {Nishio}}, \bibinfo {author}
  {\bibfnamefont {Hayao}\ \bibnamefont {Kobayashi}}, \bibinfo {author}
  {\bibfnamefont {Akiko}\ \bibnamefont {Kobayashi}}, \ and\ \bibinfo {author}
  {\bibfnamefont {Reizo}\ \bibnamefont {Kato}},\ }\bibfield  {title} {\enquote
  {\bibinfo {title} {{Magnetotransport Phenomena of $\alpha$-Type
  ({BEDT}-{TTF})$_2$I$_3$ under High Pressures}},}\ }\href {\doibase
  10.1143/jpsj.61.23} {\bibfield  {journal} {\bibinfo  {journal} {J. Phys. Soc.
  Jpn.}\ }\textbf {\bibinfo {volume} {61}},\ \bibinfo {pages} {23--26}
  (\bibinfo {year} {1992})}\BibitemShut {NoStop}%
\bibitem [{\citenamefont {Takahashi}(2003)}]{Takahashi2003}%
  \BibitemOpen
  \bibfield  {author} {\bibinfo {author} {\bibfnamefont {Toshihiro}\
  \bibnamefont {Takahashi}},\ }\bibfield  {title} {\enquote {\bibinfo {title}
  {{NMR} studies of charge ordering in organic conductors},}\ }\href {\doibase
  10.1016/s0379-6779(02)00404-6} {\bibfield  {journal} {\bibinfo  {journal}
  {Synth. Met.}\ }\textbf {\bibinfo {volume} {133-134}},\ \bibinfo {pages}
  {261--264} (\bibinfo {year} {2003})}\BibitemShut {NoStop}%
\bibitem [{\citenamefont {Seo}(2000)}]{Seo2000}%
  \BibitemOpen
  \bibfield  {author} {\bibinfo {author} {\bibfnamefont {H.}~\bibnamefont
  {Seo}},\ }\bibfield  {title} {\enquote {\bibinfo {title} {{Charge Ordering in
  Organic ET Compounds}},}\ }\href@noop {} {\bibfield  {journal} {\bibinfo
  {journal} {J. Phys. Soc. Jpn.}\ }\textbf {\bibinfo {volume} {69}},\ \bibinfo
  {pages} {805} (\bibinfo {year} {2000})}\BibitemShut {NoStop}%
\bibitem [{\citenamefont {Kino}\ and\ \citenamefont
  {Fukuyama}(1996)}]{Kino1996}%
  \BibitemOpen
  \bibfield  {author} {\bibinfo {author} {\bibfnamefont {Hiori}\ \bibnamefont
  {Kino}}\ and\ \bibinfo {author} {\bibfnamefont {Hidetoshi}\ \bibnamefont
  {Fukuyama}},\ }\bibfield  {title} {\enquote {\bibinfo {title} {{Phase Diagram
  of Two-Dimensional Organic Conductors: ({BEDT}-{TTF})$_2$X}},}\ }\href
  {\doibase 10.1143/jpsj.65.2158} {\bibfield  {journal} {\bibinfo  {journal}
  {J. Phys. Soc. Jpn.}\ }\textbf {\bibinfo {volume} {65}},\ \bibinfo {pages}
  {2158--2169} (\bibinfo {year} {1996})}\BibitemShut {NoStop}%
\bibitem [{\citenamefont {Katayama}\ \emph {et~al.}(2006)\citenamefont
  {Katayama}, \citenamefont {Kobayashi},\ and\ \citenamefont
  {Suzumura}}]{Katayama2006}%
  \BibitemOpen
  \bibfield  {author} {\bibinfo {author} {\bibfnamefont {S.}~\bibnamefont
  {Katayama}}, \bibinfo {author} {\bibfnamefont {A.}~\bibnamefont {Kobayashi}},
  \ and\ \bibinfo {author} {\bibfnamefont {Y.}~\bibnamefont {Suzumura}},\
  }\bibfield  {title} {\enquote {\bibinfo {title} {{Pressure-Induced Zero-Gap
  Semiconducting State in Organic Conductor $\alpha$-(BEDT-TTF)$_2$I$_3$
  Salt}},}\ }\href@noop {} {\bibfield  {journal} {\bibinfo  {journal} {J. Phys.
  Soc. Jpn.}\ }\textbf {\bibinfo {volume} {75}},\ \bibinfo {pages} {054705}
  (\bibinfo {year} {2006})}\BibitemShut {NoStop}%
\bibitem [{\citenamefont {Ishibashi}\ \emph {et~al.}(2006)\citenamefont
  {Ishibashi}, \citenamefont {Tamura}, \citenamefont {Kohyama},\ and\
  \citenamefont {Terakura}}]{Ishibashi2006}%
  \BibitemOpen
  \bibfield  {author} {\bibinfo {author} {\bibfnamefont {Shoji}\ \bibnamefont
  {Ishibashi}}, \bibinfo {author} {\bibfnamefont {Tomoyuki}\ \bibnamefont
  {Tamura}}, \bibinfo {author} {\bibfnamefont {Masanori}\ \bibnamefont
  {Kohyama}}, \ and\ \bibinfo {author} {\bibfnamefont {Kiyoyuki}\ \bibnamefont
  {Terakura}},\ }\bibfield  {title} {\enquote {\bibinfo {title} {{{\it Ab
  Initio} Electronic-Structure Calculations for
  $\alpha$-({BEDT}-{TTF})$_2$I$_3$}},}\ }\href {\doibase
  10.1143/jpsj.75.015005} {\bibfield  {journal} {\bibinfo  {journal} {J. Phys.
  Soc. Jpn.}\ }\textbf {\bibinfo {volume} {75}},\ \bibinfo {pages} {015005}
  (\bibinfo {year} {2006})}\BibitemShut {NoStop}%
\bibitem [{\citenamefont {Kino}\ and\ \citenamefont
  {Miyazaki}(2006)}]{Kino2006}%
  \BibitemOpen
  \bibfield  {author} {\bibinfo {author} {\bibfnamefont {Hiori}\ \bibnamefont
  {Kino}}\ and\ \bibinfo {author} {\bibfnamefont {Tsuyoshi}\ \bibnamefont
  {Miyazaki}},\ }\bibfield  {title} {\enquote {\bibinfo {title}
  {{First-Principles Study of Electronic Structure in
  $\alpha$-({BEDT}-{TTF})$_2$I$_3$ at Ambient Pressure and with Uniaxial
  Strain}},}\ }\href {\doibase 10.1143/jpsj.75.034704} {\bibfield  {journal}
  {\bibinfo  {journal} {J. Phys. Soc. Jpn.}\ }\textbf {\bibinfo {volume}
  {75}},\ \bibinfo {pages} {034704} (\bibinfo {year} {2006})}\BibitemShut
  {NoStop}%
\bibitem [{\citenamefont {Osada}(2008)}]{Osada2008}%
  \BibitemOpen
  \bibfield  {author} {\bibinfo {author} {\bibfnamefont {T.}~\bibnamefont
  {Osada}},\ }\bibfield  {title} {\enquote {\bibinfo {title} {{Negative
  Interlayer Magnetoresistance and Zero-Mode Landau Level in Multilayer Dirac
  Electron Systems}},}\ }\href@noop {} {\bibfield  {journal} {\bibinfo
  {journal} {J. Phys. Soc. Jpn.}\ }\textbf {\bibinfo {volume} {77}},\ \bibinfo
  {pages} {084711} (\bibinfo {year} {2008})}\BibitemShut {NoStop}%
\bibitem [{\citenamefont {Tajima}\ \emph {et~al.}(2009)\citenamefont {Tajima},
  \citenamefont {Sugawara}, \citenamefont {Kato}, \citenamefont {Nishio},\ and\
  \citenamefont {Kajita}}]{Tajima2009}%
  \BibitemOpen
  \bibfield  {author} {\bibinfo {author} {\bibfnamefont {N.}~\bibnamefont
  {Tajima}}, \bibinfo {author} {\bibfnamefont {S.}~\bibnamefont {Sugawara}},
  \bibinfo {author} {\bibfnamefont {R.}~\bibnamefont {Kato}}, \bibinfo {author}
  {\bibfnamefont {Y.}~\bibnamefont {Nishio}}, \ and\ \bibinfo {author}
  {\bibfnamefont {K}~\bibnamefont {Kajita}},\ }\bibfield  {title} {\enquote
  {\bibinfo {title} {{Effects of Zero-Mode Landau Level on Inter-Layer
  Magnetoresistance in Multilayer Massless Dirac Fermions System}},}\
  }\href@noop {} {\bibfield  {journal} {\bibinfo  {journal} {Phys. Rev. Lett.}\
  }\textbf {\bibinfo {volume} {102}},\ \bibinfo {pages} {176403} (\bibinfo
  {year} {2009})}\BibitemShut {NoStop}%
\bibitem [{\citenamefont {Morinari}\ \emph {et~al.}(2009)\citenamefont
  {Morinari}, \citenamefont {Himura},\ and\ \citenamefont
  {Tohyama}}]{Morinari2009}%
  \BibitemOpen
  \bibfield  {author} {\bibinfo {author} {\bibfnamefont {Takao}\ \bibnamefont
  {Morinari}}, \bibinfo {author} {\bibfnamefont {Takahiro}\ \bibnamefont
  {Himura}}, \ and\ \bibinfo {author} {\bibfnamefont {Takami}\ \bibnamefont
  {Tohyama}},\ }\bibfield  {title} {\enquote {\bibinfo {title} {{Possible
  Verification of Tilted Anisotropic Dirac Cone in
  $\alpha$-({BEDT}-{TTF})$_2$I$_3$ Using Interlayer Magnetoresistance}},}\
  }\href {\doibase 10.1143/jpsj.78.023704} {\bibfield  {journal} {\bibinfo
  {journal} {J. Phys. Soc. Jpn.}\ }\textbf {\bibinfo {volume} {78}},\ \bibinfo
  {pages} {023704} (\bibinfo {year} {2009})}\BibitemShut {NoStop}%
\bibitem [{\citenamefont {Goerbig}\ \emph {et~al.}(2008)\citenamefont
  {Goerbig}, \citenamefont {Fuchs}, \citenamefont {Montambaux},\ and\
  \citenamefont {Piechon}}]{Goerbig2008}%
  \BibitemOpen
  \bibfield  {author} {\bibinfo {author} {\bibfnamefont {M.~O.}\ \bibnamefont
  {Goerbig}}, \bibinfo {author} {\bibfnamefont {J.-N.}\ \bibnamefont {Fuchs}},
  \bibinfo {author} {\bibfnamefont {G.}~\bibnamefont {Montambaux}}, \ and\
  \bibinfo {author} {\bibfnamefont {F.}~\bibnamefont {Piechon}},\ }\bibfield
  {title} {\enquote {\bibinfo {title} {{Tilted anisotropic Dirac cones in
  quinoid-type graphene and $\alpha$-(BEDT-TTF)$_2$I$_3$}},}\ }\href@noop {}
  {\bibfield  {journal} {\bibinfo  {journal} {Phys. Rev. B}\ }\textbf {\bibinfo
  {volume} {78}},\ \bibinfo {eid} {045415} (\bibinfo {year}
  {2008})}\BibitemShut {NoStop}%
\bibitem [{\citenamefont {Tajima}\ \emph {et~al.}(2013)\citenamefont {Tajima},
  \citenamefont {Yamauchi}, \citenamefont {Yamaguchi}, \citenamefont {Suda},
  \citenamefont {Kawasugi}, \citenamefont {Yamamoto}, \citenamefont {Kato},
  \citenamefont {Nishio},\ and\ \citenamefont {Kajita}}]{Tajima2013}%
  \BibitemOpen
  \bibfield  {author} {\bibinfo {author} {\bibfnamefont {Naoya}\ \bibnamefont
  {Tajima}}, \bibinfo {author} {\bibfnamefont {Takahiro}\ \bibnamefont
  {Yamauchi}}, \bibinfo {author} {\bibfnamefont {Tatsuya}\ \bibnamefont
  {Yamaguchi}}, \bibinfo {author} {\bibfnamefont {Masayuki}\ \bibnamefont
  {Suda}}, \bibinfo {author} {\bibfnamefont {Yoshitaka}\ \bibnamefont
  {Kawasugi}}, \bibinfo {author} {\bibfnamefont {Hiroshi~M.}\ \bibnamefont
  {Yamamoto}}, \bibinfo {author} {\bibfnamefont {Reizo}\ \bibnamefont {Kato}},
  \bibinfo {author} {\bibfnamefont {Yutaka}\ \bibnamefont {Nishio}}, \ and\
  \bibinfo {author} {\bibfnamefont {Koji}\ \bibnamefont {Kajita}},\ }\bibfield
  {title} {\enquote {\bibinfo {title} {{Quantum Hall effect in multilayered
  massless Dirac fermion systems with tilted cones}},}\ }\href {\doibase
  10.1103/PhysRevB.88.075315} {\bibfield  {journal} {\bibinfo  {journal} {Phys.
  Rev. B}\ }\textbf {\bibinfo {volume} {88}},\ \bibinfo {pages} {075315}
  (\bibinfo {year} {2013})}\BibitemShut {NoStop}%
\bibitem [{\citenamefont {Moroto}\ \emph {et~al.}(2004)\citenamefont {Moroto},
  \citenamefont {Hiraki}, \citenamefont {Takano}, \citenamefont {Kubo},
  \citenamefont {Takahashi}, \citenamefont {Yamamoto},\ and\ \citenamefont
  {Nakamura}}]{Moroto2004}%
  \BibitemOpen
  \bibfield  {author} {\bibinfo {author} {\bibfnamefont {S.}~\bibnamefont
  {Moroto}}, \bibinfo {author} {\bibfnamefont {K.-I.}\ \bibnamefont {Hiraki}},
  \bibinfo {author} {\bibfnamefont {Y.}~\bibnamefont {Takano}}, \bibinfo
  {author} {\bibfnamefont {Y.}~\bibnamefont {Kubo}}, \bibinfo {author}
  {\bibfnamefont {T.}~\bibnamefont {Takahashi}}, \bibinfo {author}
  {\bibfnamefont {H.~M.}\ \bibnamefont {Yamamoto}}, \ and\ \bibinfo {author}
  {\bibfnamefont {T.}~\bibnamefont {Nakamura}},\ }\bibfield  {title} {\enquote
  {\bibinfo {title} {{Charge disproportionation in the metallic state of
  $\alpha$-(BEDT-TTF)$_2$I$_3$}},}\ }\href@noop {} {\bibfield  {journal}
  {\bibinfo  {journal} {J. Phys. IV France}\ }\textbf {\bibinfo {volume}
  {114}},\ \bibinfo {pages} {339--340} (\bibinfo {year} {2004})}\BibitemShut
  {NoStop}%
\bibitem [{\citenamefont {Kobayashi}\ \emph {et~al.}(2007)\citenamefont
  {Kobayashi}, \citenamefont {Katayama}, \citenamefont {Suzumura},\ and\
  \citenamefont {Fukuyama}}]{Kobayashi2007}%
  \BibitemOpen
  \bibfield  {author} {\bibinfo {author} {\bibfnamefont {A.}~\bibnamefont
  {Kobayashi}}, \bibinfo {author} {\bibfnamefont {S.}~\bibnamefont {Katayama}},
  \bibinfo {author} {\bibfnamefont {Y.}~\bibnamefont {Suzumura}}, \ and\
  \bibinfo {author} {\bibfnamefont {H.}~\bibnamefont {Fukuyama}},\ }\bibfield
  {title} {\enquote {\bibinfo {title} {{Massless Fermions in Organic
  Conductor}},}\ }\href@noop {} {\bibfield  {journal} {\bibinfo  {journal} {J.
  Phys. Soc. Jpn.}\ }\textbf {\bibinfo {volume} {76}},\ \bibinfo {pages}
  {034711} (\bibinfo {year} {2007})}\BibitemShut {NoStop}%
\bibitem [{\citenamefont {Sasaki}\ and\ \citenamefont
  {Morinari}(2014)}]{Sasaki2014}%
  \BibitemOpen
  \bibfield  {author} {\bibinfo {author} {\bibfnamefont {Kazuko}\ \bibnamefont
  {Sasaki}}\ and\ \bibinfo {author} {\bibfnamefont {Takao}\ \bibnamefont
  {Morinari}},\ }\bibfield  {title} {\enquote {\bibinfo {title} {{Dirac Fermion
  State with Real Space $\pi$-Flux on Anisotropic Square Lattice and Triangular
  Lattice}},}\ }\href {\doibase 10.7566/jpsj.83.034712} {\bibfield  {journal}
  {\bibinfo  {journal} {J. Phys. Soc. Jpn.}\ }\textbf {\bibinfo {volume}
  {83}},\ \bibinfo {pages} {034712} (\bibinfo {year} {2014})}\BibitemShut
  {NoStop}%
\bibitem [{\citenamefont {Morinari}\ and\ \citenamefont
  {Suzumura}(2014)}]{Morinari2014}%
  \BibitemOpen
  \bibfield  {author} {\bibinfo {author} {\bibfnamefont {Takao}\ \bibnamefont
  {Morinari}}\ and\ \bibinfo {author} {\bibfnamefont {Yoshikazu}\ \bibnamefont
  {Suzumura}},\ }\bibfield  {title} {\enquote {\bibinfo {title} {{On the
  Possible Zero-Gap State in Organic Conductor $\alpha$-({BEDT}-{TSF})$_2$I$_3$
  under Pressure}},}\ }\href {\doibase 10.7566/jpsj.83.094701} {\bibfield
  {journal} {\bibinfo  {journal} {J. Phys. Soc. Jpn.}\ }\textbf {\bibinfo
  {volume} {83}},\ \bibinfo {pages} {094701} (\bibinfo {year}
  {2014})}\BibitemShut {NoStop}%
\bibitem [{\citenamefont {Asano}\ and\ \citenamefont
  {Hotta}(2011)}]{AsanoHotta2011}%
  \BibitemOpen
  \bibfield  {author} {\bibinfo {author} {\bibfnamefont {Kenichi}\ \bibnamefont
  {Asano}}\ and\ \bibinfo {author} {\bibfnamefont {Chisa}\ \bibnamefont
  {Hotta}},\ }\bibfield  {title} {\enquote {\bibinfo {title} {{Designing Dirac
  points in two-dimensional lattices}},}\ }\href {\doibase
  10.1103/PhysRevB.83.245125} {\bibfield  {journal} {\bibinfo  {journal} {Phys.
  Rev. B}\ }\textbf {\bibinfo {volume} {83}},\ \bibinfo {pages} {245125}
  (\bibinfo {year} {2011})}\BibitemShut {NoStop}%
\bibitem [{\citenamefont {Pi{\'{e}}chon}\ and\ \citenamefont
  {Suzumura}(2013)}]{Piechon2013}%
  \BibitemOpen
  \bibfield  {author} {\bibinfo {author} {\bibfnamefont {Fr{\'{e}}d{\'{e}}ric}\
  \bibnamefont {Pi{\'{e}}chon}}\ and\ \bibinfo {author} {\bibfnamefont
  {Yoshikazu}\ \bibnamefont {Suzumura}},\ }\bibfield  {title} {\enquote
  {\bibinfo {title} {{Dirac Electron in Organic Conductor
  $\alpha$-({BEDT}-{TTF})$_2$I$_3$ with Inversion Symmetry}},}\ }\href
  {\doibase 10.7566/jpsj.82.033703} {\bibfield  {journal} {\bibinfo  {journal}
  {J. Phys. Soc. Jpn.}\ }\textbf {\bibinfo {volume} {82}},\ \bibinfo {pages}
  {033703} (\bibinfo {year} {2013})}\BibitemShut {NoStop}%
\bibitem [{Note1()}]{Note1}%
  \BibitemOpen
  \bibinfo {note} {See Supplemental Material for details about the move of the
  Dirac points in the BZ and the change of the electronic
  correlation.}\BibitemShut {Stop}%
\bibitem [{Note2()}]{Note2}%
  \BibitemOpen
  \bibinfo {note} {See Supplemental Material for details about symmetry
  breaking}\BibitemShut {NoStop}%
\bibitem [{\citenamefont {McKenzie}\ and\ \citenamefont
  {Moses}(1998)}]{McKenzie1998}%
  \BibitemOpen
  \bibfield  {author} {\bibinfo {author} {\bibfnamefont {Ross~H.}\ \bibnamefont
  {McKenzie}}\ and\ \bibinfo {author} {\bibfnamefont {Perez}\ \bibnamefont
  {Moses}},\ }\bibfield  {title} {\enquote {\bibinfo {title} {{Incoherent
  Interlayer Transport and Angular-Dependent Magnetoresistance Oscillations in
  Layered Metals}},}\ }\href {\doibase 10.1103/physrevlett.81.4492} {\bibfield
  {journal} {\bibinfo  {journal} {Phys. Rev. Lett.}\ }\textbf {\bibinfo
  {volume} {81}},\ \bibinfo {pages} {4492--4495} (\bibinfo {year}
  {1998})}\BibitemShut {NoStop}%
\bibitem [{\citenamefont {Sugawara}\ \emph {et~al.}(2010)\citenamefont
  {Sugawara}, \citenamefont {Tamura}, \citenamefont {Tajima}, \citenamefont
  {Kato}, \citenamefont {Sato}, \citenamefont {Nishio},\ and\ \citenamefont
  {Kajita}}]{Sugawara2010mr}%
  \BibitemOpen
  \bibfield  {author} {\bibinfo {author} {\bibfnamefont {Shigeharu}\
  \bibnamefont {Sugawara}}, \bibinfo {author} {\bibfnamefont {Masafumi}\
  \bibnamefont {Tamura}}, \bibinfo {author} {\bibfnamefont {Naoya}\
  \bibnamefont {Tajima}}, \bibinfo {author} {\bibfnamefont {Reizo}\
  \bibnamefont {Kato}}, \bibinfo {author} {\bibfnamefont {Mitsuyuki}\
  \bibnamefont {Sato}}, \bibinfo {author} {\bibfnamefont {Yutaka}\ \bibnamefont
  {Nishio}}, \ and\ \bibinfo {author} {\bibfnamefont {Koji}\ \bibnamefont
  {Kajita}},\ }\bibfield  {title} {\enquote {\bibinfo {title} {{Temperature
  Dependence of Inter-Layer Longitudinal Magnetoresistance in
  $\alpha$-(BEDT-TTF)$_{2}$I$_{3}$: Positive versus Negative Contributions in a
  Tilted Dirac Cone System}},}\ }\href {\doibase 10.1143/JPSJ.79.113704}
  {\bibfield  {journal} {\bibinfo  {journal} {J. Phys. Soc. Jpn.}\ }\textbf
  {\bibinfo {volume} {79}},\ \bibinfo {pages} {113704} (\bibinfo {year}
  {2010})}\BibitemShut {NoStop}%
\bibitem [{\citenamefont {Tajima}\ and\ \citenamefont
  {Morinari}(2018)}]{Tajima2018}%
  \BibitemOpen
  \bibfield  {author} {\bibinfo {author} {\bibfnamefont {Naoya}\ \bibnamefont
  {Tajima}}\ and\ \bibinfo {author} {\bibfnamefont {Takao}\ \bibnamefont
  {Morinari}},\ }\bibfield  {title} {\enquote {\bibinfo {title} {{Tilted Dirac
  Cone Effect on Interlayer Magnetoresistance in
  $\alpha$-({BEDT}-{TTF})$_2$I$_3$}},}\ }\href {\doibase
  10.7566/jpsj.87.045002} {\bibfield  {journal} {\bibinfo  {journal} {J. Phys.
  Soc. Jpn.}\ }\textbf {\bibinfo {volume} {87}},\ \bibinfo {pages} {045002}
  (\bibinfo {year} {2018})}\BibitemShut {NoStop}%
\bibitem [{\citenamefont {Winter}\ \emph {et~al.}(2017)\citenamefont {Winter},
  \citenamefont {Riedl},\ and\ \citenamefont {Valent{\'{\i}}}}]{Winter2017}%
  \BibitemOpen
  \bibfield  {author} {\bibinfo {author} {\bibfnamefont {Stephen~M.}\
  \bibnamefont {Winter}}, \bibinfo {author} {\bibfnamefont {Kira}\ \bibnamefont
  {Riedl}}, \ and\ \bibinfo {author} {\bibfnamefont {Roser}\ \bibnamefont
  {Valent{\'{\i}}}},\ }\bibfield  {title} {\enquote {\bibinfo {title}
  {{Importance of spin-orbit coupling in layered organic salts}},}\ }\href
  {\doibase 10.1103/physrevb.95.060404} {\bibfield  {journal} {\bibinfo
  {journal} {Phys. Rev. B}\ }\textbf {\bibinfo {volume} {95}},\ \bibinfo
  {pages} {060404} (\bibinfo {year} {2017})}\BibitemShut {NoStop}%
\bibitem [{\citenamefont {Osada}(2018)}]{Osada2018}%
  \BibitemOpen
  \bibfield  {author} {\bibinfo {author} {\bibfnamefont {Toshihito}\
  \bibnamefont {Osada}},\ }\bibfield  {title} {\enquote {\bibinfo {title}
  {{Topological Insulator State due to Finite Spin{\textendash}Orbit
  Interaction in an Organic Dirac Fermion System}},}\ }\href {\doibase
  10.7566/jpsj.87.075002} {\bibfield  {journal} {\bibinfo  {journal} {J. Phys.
  Soc. Jpn.}\ }\textbf {\bibinfo {volume} {87}},\ \bibinfo {pages} {075002}
  (\bibinfo {year} {2018})}\BibitemShut {NoStop}%
\bibitem [{Note3()}]{Note3}%
  \BibitemOpen
  \bibinfo {note} {See Supplemental Material for details about the effect of
  spin-orbit coupling.}\BibitemShut {Stop}%
\end{thebibliography}%

\end{document}


\title{
Supplementary Material
}



\maketitle

\setcounter{section}{0}
\setcounter{page}{1}
\setcounter{equation}{0}
\setcounter{figure}{0}
\renewcommand{\thepage}{S\arabic{page}}  
\renewcommand{\thesection}{S\arabic{section}}
\renewcommand{\theequation}{S\arabic{equation}}
\renewcommand{\thefigure}{S\arabic{figure}}
\renewcommand{\thetable}{S\arabic{table}} 

\section{
  Explicit form of the mean field Hamiltonian
}
The mean field Hamiltonian
(Eq.~(5) in the main text) is written as
\bea
    {{\mathcal H}_{\textrm{mf}}}
    &=& \sum_{\bm{k},\sigma,\sigma'} {c_{\bm{k}\sigma}^\dag
    \left[
      {\mathcal H}\left( {\bm{k}} \right)
      \right]_{\sigma \sigma'}
         {c_{\bm{k}\sigma'}}}
    \nonumber \\
&=& \sum\limits_{{\bm{k}},\alpha ,\beta ,\sigma ,\sigma '} {c_{{\bm{k}}\alpha \sigma }^\dag {{\left[ {{\mathcal H}\left( {\bm{k}} \right)} \right]}_{\alpha \sigma ,\beta \sigma '}}{c_{{\bm{k}}\beta \sigma '}}} 
\eea
The matrix elements are given by
\bea
    \left[
      {\mathcal H}\left( {\bm{k}} \right)
      \right]_{\alpha \sigma, \beta \sigma'}
    &=&
    {{\left[ {H_{\bm{k}}^{\left(  +  \right)}} \right]}
        _{\alpha \sigma ,\beta \sigma '}}
    {e^{i{\bm{k}} \cdot {\bm{d}}_{\alpha \beta }^{\left(  +  \right)}}}
    \nonumber \\
    & & + {{\left[ {H_{\bm{k}}^{\left(  -  \right)}} \right]}
        _{\alpha \sigma ,\beta \sigma '}}
      {e^{ - i{\bm{k}} \cdot {\bm{d}}_{\alpha \beta }^{\left(  -  \right)}}}
      \nonumber \\  & &  \left.
      + H_{{\bm{k}}\alpha \sigma }^{\left( c \right)}
                {\delta _{\alpha \beta }}{\delta _{\sigma \sigma '}}
                \right],
                  \label{supp:eqH2D}
    \eea
where
\be
{\left[ {H_{\bm{k}}^{\left(  \pm  \right)}} \right]_{\alpha \sigma ,\beta \sigma '}} = t_{\alpha \beta }^{\left(  \pm  \right)} - {V_{\alpha \beta }}\chi _{\alpha \sigma ,\beta \sigma ', \pm }^*   
\ee
with $t_{13}^{\left(  +  \right)} = {t_{b3}}$,
$t_{13}^{\left(  -  \right)} = {t_{b2}}$, etc.
The on-site term is given by
\be
H_{{\bm{k}}\alpha \sigma }^{\left( c \right)} = 2\sum\limits_\gamma  {{V_{\alpha \gamma }}{n_\gamma }}  + U\left\langle {{n_{\alpha \overline \sigma  }}} \right\rangle,
\ee
where $\overline \sigma$ is flipped to $\sigma$.

In our self-consistent calculation,
we diagonalize this mean field Hamiltonian
at each momentum.
We take $200 \times 200$ for
the number of the BZ points.
${\chi _{\alpha \sigma ,\beta \sigma ', \pm }}$
and
${{n_{\alpha \sigma }}}$ values
are computed by calculating the expectation values,
$\left\langle c_{{\bm{k}}\alpha \sigma }^\dagger c_{{\bm{k}}\beta \sigma '} \right\rangle$.

\section{
  \label{supp:symm}
  Symmetries of the Hamiltonian at the Dirac points
  }
As discussed in the main text,
the exchange correlation breaks
both the TRS and the inversion symmetry.
However, some components of the Hamiltonians at the Dirac points
retain the inversion symmetry.
To clarify this point,
we represent the Hamiltonian
at each Dirac point
in terms of ${\mathcal X}_{\mu\nu\lambda}$,
where the coefficients are as shown in
Table \ref{supp:table1}.
\begin{table}[b]
  \caption{\label{supp:table1}
    Decomposition of the Hamiltonian at each Dirac point
    into ${\mathcal X}_{\mu\nu\lambda}$.
    For the case of $U=0.4$ and $V_c=V_p=0$,
    both the TRS and inversion symmetry are unbroken.
    The Dirac points are at ${\bm k}={\bm k}_D, -{\bm k}_D$,
    with ${\bm k}_D=\left(1.5516, -0.8044 \right)$.
    Meanwhile,
    for the case of $U=0.4$, $V_c=0.17$, and $V_p=0.05$,
    these symmetries are broken,
    and
    the Dirac points are at ${\bm k}={\bm k}_D^{(1)}, {\bm k}_D^{(2)}$,
    with ${\bm k}_D^{(1)} \neq -{\bm k}_D^{(2)}$,
    ${\bm k}_D^{(1)}=\left(0.9296,-0.8009\right)$, and
    ${\bm k}_D^{(2)}=\left(-0.9018,0.7618\right)$.
    The breaking of the TRS and the inversion symmetry
    is clearly illustrated through the comparison of these two cases.
}
\begin{ruledtabular}
  \begin{tabular}{lrrrr}
    ${\mathcal X}_{\mu\nu\lambda}$ &
    \multicolumn{1}{c}{${\bm{k}_D}$} &
    \multicolumn{1}{c}{$-{\bm{k}_D}$} &
    \multicolumn{1}{c}{${\bm{k}_D^{(1)}}$} &
    \multicolumn{1}{c}{${\bm{k}_D^{(2)}}$} \\
    \colrule
${\mathcal X}_{010}$  & 0.0000387   & 0.0000387  & 0.0097423   & -0.0092337  \\
${\mathcal X}_{020}$  & 0.0610626   & -0.0610626  & 0.0617559   & -0.0603957  \\
${\mathcal X}_{030}$  & -0.0102378   & -0.0102378  & 0.0227222   & 0.0227222  \\
${\mathcal X}_{100}$  & -0.0687389   & -0.0687389  & -0.0237383   & -0.0105359  \\
${\mathcal X}_{110}$  & -0.0687389   & -0.0687389  & -0.0237383   & -0.0105359  \\
${\mathcal X}_{120}$  & -0.1041172   & 0.1041172  & -0.1511936   & 0.1524187  \\
${\mathcal X}_{130}$  & -0.0650720   & -0.0650720  & -0.1106209   & -0.1070677  \\
${\mathcal X}_{200}$  & 0.0173256   & -0.0173256  & -0.0398026   & 0.0484009  \\
${\mathcal X}_{210}$  & -0.0173256   & 0.0173256  & 0.0398026   & -0.0484009  \\
${\mathcal X}_{220}$  & -0.0650720   & -0.0650720  & -0.1106209   & -0.1070677  \\
${\mathcal X}_{230}$  & -0.1041172   & 0.1041172  & -0.1511936   & 0.1524187  \\
${\mathcal X}_{300}$  & -0.0160446   & -0.0160446  & -0.0152960   & -0.0152960  \\
${\mathcal X}_{310}$  & 0.0290397   & 0.0290397  & 0.0265949   & 0.0379016  \\
${\mathcal X}_{320}$  & 0.0610626   & -0.0610626  & 0.0617559   & -0.0603957  \\
${\mathcal X}_{330}$  & 0.0102378   & 0.0102378  & -0.0227222   & -0.0227222  \\
  \end{tabular}
\end{ruledtabular}
\end{table}
In case of $U=0.4$ and $V_c=V_p=0$,
the Dirac points are located at ${\bm k}={\bm k}_D, -{\bm k}_D$ and
both the TRS and the inversion symmetry
are unbroken.
Under these symmetries,
the Hamiltonian, in general, has the form
\be
   {\mathcal H}\left( {\bm{k}} \right) =
   \left( {\begin{array}{cccc}
       {a} &
       {x - iy} &
       {p} &
       {q} \\
       {x + iy} &
       {a} &
       {{p^*}} &
       {{q^*}}\\
       {{p^*}} &
       {p} &
       { - a + b} &
       {c}\\
       {{q^*}} &
       {q} &
       {c} &
       { - a - b}
   \end{array}} \right),
   \ee
   where $a$, $b$, $c$, $x$, and $y$
   are real functions of ${\bm{k}}$ and
   $p$ and $q$ are complex functions of ${\bm{k}}$.
   The terms $y$, ${\textrm{Im}}~p$, and
   ${\textrm{Im}}~q$ are odd,
   while the others are even.
   Based on this form,
   we see that
   some of the coefficients are the same,
   such as
   ${\mathcal X}_{02 0}$
   and
   ${\mathcal X}_{32 0}$,
   and
   some of the coefficients have opposite signs,
   such as 
   ${\mathcal X}_{20 0}$
   and
   ${\mathcal X}_{21 0}$.
   
When $U=0.4$, $V_c=0.17$, and $V_p=0.05$,
both the TRS and the inversion symmetry
are broken
but the symmetries in the coefficients are preserved.
The symmetries are broken by
the functions multiplied to ${\mathcal X}_{\mu\nu0}$
that are neither even nor odd
with respect to $\bm{k}$.
Despite the presence of such symmetry breaking factors,
part of the inversion symmetry
is still preserved.
In fact, we find
\be
{\textrm{tr}}\left[ {{{\mathcal X}_{030}}
    {\mathcal H}\left( {{\bm{k}}_D^{\left( {1,2} \right)}} \right)} \right]
=  - {\textrm{tr}}\left[ {{{\mathcal X}_{330}}
    {\mathcal H}\left( {{\bm{k}}_D^{\left( {1,2} \right)}} \right)} \right].
\ee
This is indispensable because
${\mathcal X}_{030}-{\mathcal X}_{330}$
and ${\mathcal X}_{300}$ combined with the identity matrix
lead to degenerate and other separated levels.
We note that these generators are diagonal matrices
and that diagonal components arise from the charge at each molecule.
Therefore, they are associated with the inversion symmetry
between $n_{\textrm A}$ and $n_{{\textrm A}^{\prime}}$.

\section{
  Dirac point motion in the BZ
  and the change of the electronic correlation
  }
\begin{figure}[tbp]
  \includegraphics[width=\linewidth,clip]{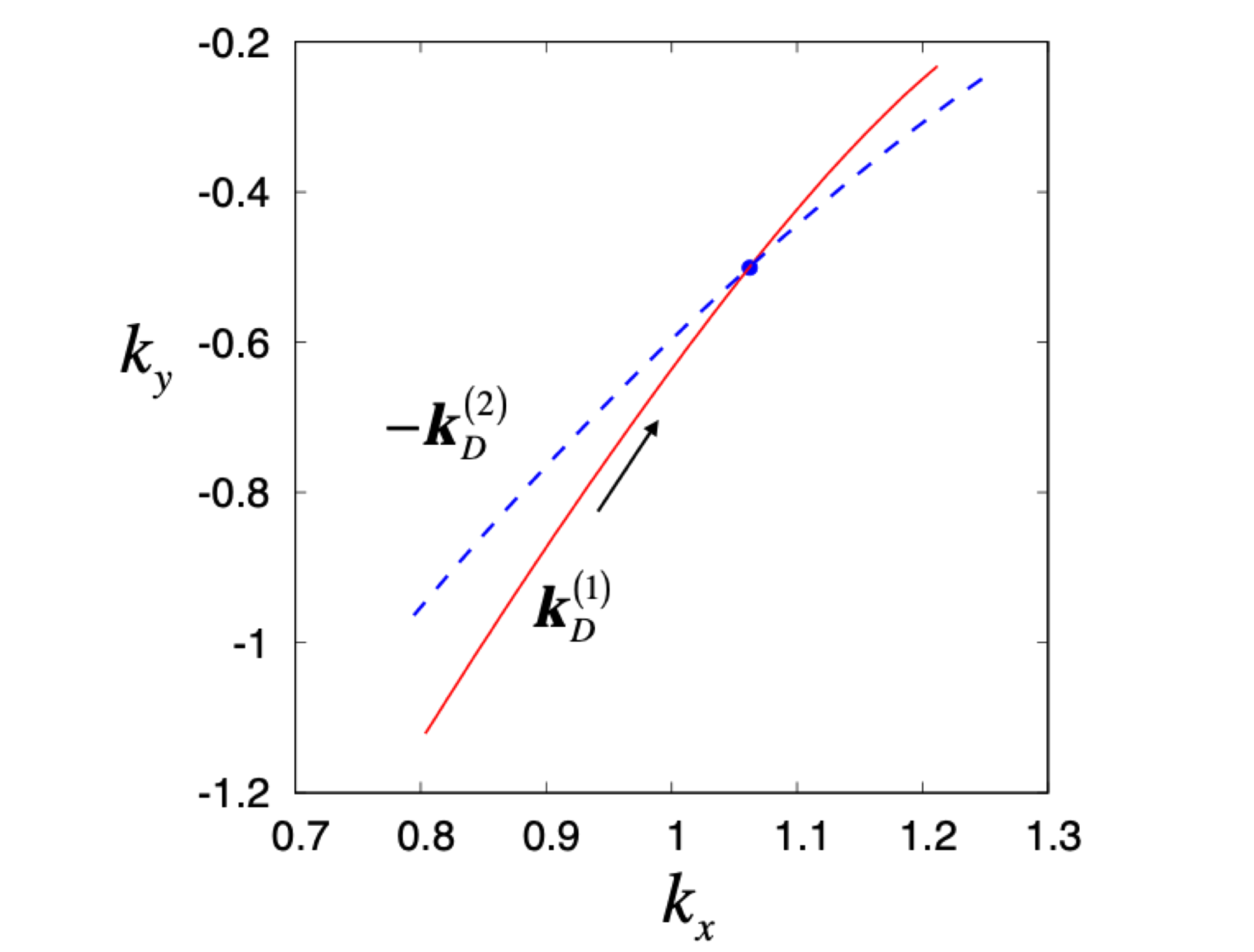}
\caption{
  \label{fig:DPmotion}
  Dirac point motion in 2D BZ.
  The two Dirac points are denoted by 
  ${\bm k}_{\textrm D}^{(1)}$
  and
  ${\bm k}_{\textrm D}^{(2)}$.
  The pressure range is $0.50 \leq P \leq 2.0$.
  The interaction parameters are 
  $U=0.4$, $V_c=0.17$, and $V_p=0.05$.
  As we increase the pressure, the two Dirac points move 
  in the BZ.
  Here, we show ${\bm k}_{\textrm D}^{(1)}$ and
  $-{\bm k}_{\textrm D}^{(2)}$.
}
\end{figure}

As discussed in the main text,
the Dirac points are not symmetrically located 
with respect to their origin
in the 2D Dirac semimetal state.
The positions of the Dirac points
move by changing the pressure,
as shown in Fig.~\ref{fig:DPmotion}.
There is an accidental TRS point,
where $k_x=1.06$ and $k_y=-0.501$ 
at $P=1.25$.

\widetext
\section{
  The Hamiltonian with three-dimensional wave vector
}
  In the presence of the inter-layer hopping,
  the Hamiltonian is given by
\be
   {\mathcal H}\left( {k_x},{k_y},{k_z} \right)
   = {\mathcal H}\left( {{k_x},{k_y}} \right) + \left( {\begin{array}{*{20}{c}}
{2{t_1}\cos {k_z}}&{2{t_2}\cos {k_z}}&0&0\\
{2{t_2}\cos {k_z}}&{2{t_1}\cos {k_z}}&0&0\\
0&0&{2{t_1}\cos {k_z}}&{2{t_2}\cos {k_z}}\\
0&0&{2{t_2}\cos {k_z}}&{2{t_1}\cos {k_z}}
   \end{array}} \right) \otimes {s_0}.
   \label{supp:eqH3D}
\ee
Here, we denote dependence of $k_x$, $k_y$, and $k_z$, explicitly.
The first term ${\mathcal H}\left( {{k_x},{k_y}} \right)$
is defined in Eq.~(\ref{supp:eqH2D}).

  As discussed in the main text, we obtain
  the linear energy dispersion
  of the three-dimensional Dirac semimetal phase
  from the diagonalization of Eq.~(\ref{supp:eqH3D}).
  However, its demonstration requires numerical calculations.
  In order to analytically illustrate the linear energy dispersion
  of the three-dimensional Dirac semimetal phase,
we consider a simple model:
\be
{\mathcal H}^{(1)}\left( {k_x},{k_y},{k_z} \right)
= \left( {\begin{array}{*{20}{c}}
{2{t_1}\cos {k_z}}&{2{t_2}\cos {k_z}}&{v\left( {{k_x} - i{k_y}} \right)}&0\\
{2{t_2}\cos {k_z}}&{2{t_1}\cos {k_z}}&0&{ - v\left( {{k_x} - i{k_y}} \right)}\\
{v\left( {{k_x} + i{k_y}} \right)}&0&{2{t_1}\cos {k_z}}&{2{t_2}\cos {k_z}}\\
0&{ - v\left( {{k_x} + i{k_y}} \right)}&{2{t_2}\cos {k_z}}&{2{t_1}\cos {k_z}}
      \end{array}} \right) \otimes {s_0} .
      \ee
      Here, $v$ is a constant.
Setting ${k_z} = \pi /2 + {\kappa _z}$,
we obtain
\be
   {E^{(1)}_{{k_x},{k_y},{k_z}}} =  \pm \sqrt {{v^2}\left( {k_x^2 + k_y^2} \right) + 4t_2^2{{\sin }^2}{\kappa _z}}  - 2{t_1}\sin {\kappa _z}
\ee
If $|\kappa_z| \ll 1$, we obtain the linear energy dispersion
of the three-dimensional Dirac semimetal phase
with the tilt in $k_z$ direction,
\be
   {E^{(1)}_{{k_x},{k_y},{k_z}}} =  \pm \sqrt {{v^2}\left( {k_x^2 + k_y^2} \right) + 4t_2^2\kappa _z^2}  - 2{t_1}{\kappa _z}.
\ee
Note that these energy dispersions with either plus sign or minus sign
are doubly degenerate along with spin degeneracy.

A more realistic model is the following Hamiltonian:
\be
{\mathcal H}^{(2)}\left( {k_x},{k_y},{k_z} \right)
= \left( {\begin{array}{*{20}{c}}
-2{t_1}{k_z}&{A + iv{k_y} - 2{t_2}{k_z}}&{B + v{k_x}}&0\\
{A - iv{k_y} - 2{t_2}{k_z}}& -2{t_1}{k_z} &0&{B - v{k_x}}\\
{B + v{k_x}}&0& -2{t_1}{k_z} &{ - A + i{k_y} - 2{t_2}{k_z}}\\
0&{B - v{k_x}}&{ - A - i{k_y} - 2{t_2}{k_z}}& -2{t_1}{k_z}
\end{array}} \right) \otimes {s_0},
  \ee
where $A$, $B$, $v$ are constants.
Here, $\cos k_z$ is replaced by $-k_z$.
The energy dispersions are obtained as follows:
\be
E_{{k_x},{k_y},{k_z}}^{\left( { s_1 , s_2 } \right)} =  s_1 \sqrt {{A^2} + {B^2} + {v^2}\left( {k_x^2 + k_y^2} \right) + 4t_2^2k_z^2 +2 s_2 \sqrt {{A^2} + {B^2}} \sqrt {{v^2}\left( {k_x^2 + \frac{{{B^2}}}{{{A^2} + {B^2}}}k_y^2} \right) + 4t_2^2k_z^2} }  - 2{t_1}{k_z},
\label{supp:eqE2}
\ee
with $s_{1,2} = \pm 1$.
One can confirm that 
this is the energy dispersion
of the three-dimensional Dirac semimetal phase
with the tilt in $k_z$ direction as follows.
We set
\be
\eta  = \frac{1}{{\sqrt {{A^2} + {B^2}} }}\sqrt {{v^2}\left( {k_x^2 + \frac{{{B^2}}}{{{A^2} + {B^2}}}k_y^2} \right) + 4t_2^2k_z^2},
\ee
and
\be
a\eta  = \frac{A}{{{A^2} + {B^2}}}v{k_y}.
\ee
Equation~(\ref{supp:eqE2}) is rewritten as
\be
\frac{{E_{{k_x},{k_y},{k_z}}^{\left( {s_1,s_2} \right)} + 2{t_1}{k_z}}}{{\sqrt {{A^2} + {B^2}} }}
=  s_1 {f_{s_2} }\left( {\eta ,a} \right),
\ee
where
\be
{f_{s} }\left( {\eta ,a} \right) = \sqrt {1 +2s \eta  + \left( {1 + {a^2}} \right){\eta ^2}}.
\ee
Now we find
\be
   {\left. {\frac{{\partial {f_{\pm} }}}{{\partial \eta }}} \right|_{\eta  \to 0}}
   =  \pm 1
\ee
Therefore,
$E_{{k_x},{k_y},{k_z}}^{\left( {s_1,s_2} \right)} + 2{t_1}{k_z}$
is linear in $\eta$ for $\eta \ll 1$.
This observation confirms that
Eq.~(\ref{supp:eqE2}) describes
the energy dispersion of 
the three-dimensional Dirac semimetal phase
with the tilt in $k_z$ direction.

When ${v^2}\left( {k_x^2 + k_y^2} \right) + 4t_2^2k_z^2 \ll {A^2} + {B^2}$,
the approximate form is obtained as follows:
\be
E_{{k_x},{k_y},{k_z}}^{\left( {s_1,s_2} \right)} \simeq
s_1\sqrt {{A^2} + {B^2}}
+ s_2 \sqrt {{v^2}\left( {k_x^2
    + \frac{{{B^2}}}{{{A^2} + {B^2}}}k_y^2} \right) + 4t_2^2k_z^2}  - 2{t_1}{k_z}.
\ee
There are four energy bands
and the upper two bands describe
the energy dispersion
of the three-dimensional Dirac semimetal phase
as in \alphaI.
However, there is an additional symmetry in this model.
The lower two bands is a copy
of the upper two bands.
The difference is just the origin of the energy.
The lower two bands are obtained by
shifting the energy of the upper two bans
by $2\sqrt {{A^2} + {B^2}}$.
There is no such symmetry in \alphaI,
though one can show that the lower two bands also have Dirac cones.

\bibliography{../../../../../refs/lib}